%% file: paper.tex
\documentclass[12pt]{article}

\usepackage[dvipsnames]{xcolor}   
\usepackage{lmodern}
\usepackage{graphicx}             
\DeclareGraphicsExtensions{.pdf}  
\usepackage{tikz}                 
\usetikzlibrary{shapes,arrows}
\usepackage{subfigure}            
\usepackage{amsmath}              
\usepackage{amssymb}              
\usepackage{amsthm}               
\usepackage[squaren,mediumqspace]{SIunits}     
\usepackage{multirow}             
\usepackage{tabularx}             
\usepackage{slashed}              
\usepackage{booktabs}             
\usepackage{import}               
\usepackage{bigdelim}             
\usepackage{dsfont}               
\usepackage{feynmp}               
\usepackage{wasysym}              
\usepackage{cite}                 
\usepackage{hyphenat}             
\usepackage{datetime}
\usepackage[utf8]{inputenc}
\usepackage{numprint}             
\usepackage{ulem}

\input{paperdef}

\usepackage[pdftitle={Letter}, pdfauthor={Peter Drechsel, Ramona Gr\"ober}, pdfkeywords={higgs mass, precision calculation, NMSSM}, bookmarks=true, pdfborder={0 0 0}]{hyperref}
\usepackage[all]{hypcap}

\evensidemargin \oddsidemargin
\begin{document}

\thispagestyle{empty}

\def\thefootnote{\fnsymbol{footnote}}

\begin{flushleft}
  CP3-Origins-2016-054 \hfill IPPP/16/122 \\
  DESY 16-244 \hfill KA-TP-43-2016 \\
  IFT-UAM/CSIC-16-139 \hfill RM3-TH/16-13
\end{flushleft}

\vspace{0.5cm}

\begin{center}
{\large\sc {\bf Higgs-Boson Masses and Mixing Matrices in the NMSSM:\\[.5em]
Analysis of On-Shell Calculations}}
\vspace{0.5cm}

\vspace{1cm}

{\sc
P.~Drechsel$^{1}$%
\footnote{email: peter.drechsel@desy.de}%
, R.~Gr\"ober$^{2}$%
\footnote{email: ramona.groeber@durham.ac.uk}%
, S.~Heinemeyer$^{3}$%
\footnote{email: Sven.Heinemeyer@cern.ch}%
, M.~M\"uhlleitner$^{4}$%
\footnote{email: milada.muehlleitner@kit.edu }%
,\\[.3em] H.~Rzehak$^{5}$%
\footnote{email: rzehak@cp3.sdu.dk}%
,~and G.~Weiglein$^{1}$%
\footnote{email: Georg.Weiglein@desy.de}
}

\vspace*{.7cm}

{\sl
$^1$DESY, Notkestra\ss e 85, D--22607 Hamburg, Germany

\vspace*{0.1cm}

$^2$ Institute for Particle Physics Phenomenology, Department of Physics,
Durham University, Durham DH1 3LE, UK \\
INFN, Sezione di Roma Tre, Via della Vasca Navale 84, 00146 Rome, Italy

\vspace*{0.1cm}

$^3$Campus of International Excellence UAM+CSIC, 
Cantoblanco, 28049 Madrid, Spain\\
Instituto de F\'isica Te\'orica, (UAM/CSIC), Universidad
   Aut\'onoma de Madrid,\\ Cantoblanco, 28049 Madrid, Spain\\
Instituto de F\'isica de Cantabria (CSIC-UC), 39005 Santander,  Spain
\vspace*{0.1cm}

$^4$ Institute for Theoretical Physics, Karlsruhe Institute of Technology, 76128 Karlsruhe, Germany

\vspace*{0.1cm}

$^5$ $\text{CP}^3$-Origins,
University of Southern Denmark,
Campusvej 55,
5230 Odense M,
Denmark


}

\end{center}

\vspace*{0.1cm}

\begin{abstract}
\noindent
We analyze  the Higgs-boson  masses and mixing  matrices in  the NMSSM
based  on an  on-shell  (OS) renormalization  of  the gauge-boson  and
Higgs-boson masses and the parameters of the top/scalar top sector. We
compare  the  implementation  of  the OS  calculations  in  the  codes
\NC\  and \NFH\  up to  \order{\alt\als}. We  identify the  sources of
discrepancies  at the  one- and  at  the two-loop  level.  Finally  we
compare  the OS  and \DRbar\  evaluation as  implemented in  \NC.  The
results are important  ingredients for an estimate  of the theoretical
precision of Higgs-boson mass calculations in the NMSSM.
\end{abstract}

\def\thefootnote{\arabic{footnote}}
\setcounter{page}{0}
\setcounter{footnote}{0}

\newpage



\input{sections/introduction}
\input{sections/NMSSM}
\input{sections/codes}

\input{sections/scenarios}

\input{sections/calculation}

\input{sections/calculationNC}

\input{sections/conclusion}

\section*{Acknowledgements}

We acknowledge useful discussions with Kathrin Walz and Dao Thi Nhung.
H.R.'s work  is partially  supported by  the Danish  National Research
Foundation under  grant DNRF:90.   The work of  S.H.\ is  supported in
part by  CICYT (grant  FPA 2013-40715-P) and  by the  Spanish MICINN's
Consolider-Ingenio 2010  Program under grant  MultiDark CSD2009-00064.
The work of G.W.\ is supported in  part by the DFG through the SFB~676
``Particles,  Strings and  the Early  Universe'' and  by the  European
Commission  through   the  ``HiggsTools''  Initial   Training  Network
PITN-GA-2012-316704.   R.G.    is  supported   by  a   European  Union
COFUND/Durham  Junior Research  Fellowship under  the EU  grant number
609412.

\pagestyle{plain}             

\bibliographystyle{h-physrev}     
\bibliography{literature}{}


\end{document}

%% file: paperdef.tex
\newcommand{\mathi}{\mathrm{i}}

    \newcommand{\UY}{\ensuremath{U(1)_{Y}}}

\newcommand{\SUL}{\ensuremath{SU(2)_\text{L}}}

\newcommand{\SUc}{\ensuremath{SU(3)_\text{c}}}

\newcommand{\FH}{\texttt{Feyn\-Higgs}}

\newcommand{\NFH}{\texttt{NMSSM\hyp{}Feyn\-Higgs}}
\newcommand{\NFHs}{\texttt{N\hyp{}FH}}
\newcommand{\NC}{\texttt{NMSSM\-CALC}}
\newcommand{\NCs}{\texttt{NC}}
\newcommand{\MTO}{\ensuremath{m_t}}

\newcommand{\CP}{\ensuremath{{\cal CP}}}

\newcommand{\mueff}{\ensuremath{\mu_{\text{eff}}}}

\newcommand{\Al}{\ensuremath{A_\lambda}}

\newcommand{\TB}{\ensuremath{\tan{\beta}}}

\newcommand\mstop[1]{m_{\tilde{t}_{#1}}}

\newcommand\msqd{M_{\tilde{Q}_3}}

\theoremstyle{definition}

\newcommand{\order}[1]{\ensuremath{{\cal O}(#1)}}
\newcommand{\cp}{\ensuremath{{\cal CP}}}
\newcommand{\al}{\alpha}
\newcommand{\als}{\al_s}
\newcommand{\alt}{\al_t}
\newcommand{\alb}{\al_b}
\newcommand{\be}{\beta}
\newcommand{\tb}{\tan\be}

\newcommand{\gev}{\giga\electronvolt}
\newcommand{\mev}{\mega\electronvolt}

\newcommand\MW{M_W}
\newcommand\MZ{M_Z}

\newcommand\MHp{\ensuremath{M_{H^\pm}}}

\newcommand{\mt}{m_t}

\newcommand{\At}{A_t}
\newcommand{\yt}{Y_t}
\newcommand{\Ala}{A_\lambda}

\newcommand{\Ak}{A_k}

\newcommand{\MS}{\ensuremath{M_S}}

\newcommand{\DRbar}{\ensuremath{\overline{\mathrm{DR}}}}
\newcommand{\MSbar}{\ensuremath{\overline{\mathrm{MS}}}}
\newcommand{\OS}{\ensuremath{\mathrm{OS}}}

\newcommand{\ZH}[1]{\ensuremath{U^{h}_{#1}}}
\newcommand{\ZHabs}[1]{\ensuremath{|U^{h}_{#1}|}}

\def\refeq#1{\mbox{eq.~(\ref{#1})}}

\def\reffi#1{\mbox{fig.~\ref{#1}}}

\def\refta#1{\mbox{tab.~\ref{#1}}}

\def\refse#1{\mbox{sect.~\ref{#1}}}

\def\citere#1{\mbox{ref.~\cite{#1}}}

%% file: sections/introduction.tex
\section{Introduction}
\label{sec:Introduction}

The  experimental   value  of  the   mass  of  the   discovered  Higgs
boson~\cite{CMS:2015kwa},
\begin{align}
 m_H = 125.09 \pm 0.21(\rm{stat.}) \pm 0.11(\rm{syst.})~\giga\electronvolt,
\end{align}
\noindent
has an  uncertainty of  just a  few permille so  that only  four years
after  the  discovery of  the  Higgs  boson  its  mass has  become  an
electroweak precision observable.   In order to make full  use of this
high-accuracy measurement each prediction  for this quantity should be
provided  with  a  similar  precision.  Furthermore,  for  a  reliable
calculation of  the Higgs-boson mass it  is important to make  a solid
estimate for the theoretical  uncertainty of the available prediction.
Two  different  sources for  theoretical  uncertainties  exist in  the
Higgs-boson mass predictions.   One is due to  the experimental errors
of   the   Standard   Model  (SM)   input   parameters   (``parametric
uncertainties''), the  other are  unknown higher-order  corrections in
the Higgs-boson mass calculation itself (``intrinsic uncertainties'').

Supersymmetry  (SUSY)  is one  of  the  most attractive  solutions  to
several shortcomings of  the SM.  It can solve  the hierarchy problem,
provides a  Dark Matter candidate  and leads  to a unification  of the
gauge couplings,  thus paving the way  to a Grand Unified  Theory. The
most  frequently   studied  realizations  of  SUSY   are  the  Minimal
Supersymmetric     Standard      Model     (MSSM)~\cite{Gunion:1989we,
  Martin:1997ns,     Dawson:1997tz,     Djouadi:2005gj}    and     the
Next-to-Minimal         Supersymmetric          Standard         Model
(NMSSM)~\cite{Fayet:1974pd,Barbieri:1982eh,               Dine:1981rt,
  Nilles:1982dy,   Frere:1983ag,   Derendinger:1983bz,   Ellis:1988er,
  Drees:1988fc, Ellwanger:1993xa,  Ellwanger:1995ru, Ellwanger:1996gw,
  Elliott:1994ht,    King:1995vk,   Franke:1995tc,    Maniatis:2009re,
  Ellwanger:2009dp}.  In  contrast to  the SM, in  the MSSM  two Higgs
doublets are  required.  After  electroweak symmetry  breaking (EWSB),
this results in  five physical Higgs bosons.   In the $\CP$-conserving
case,  these are  two  $\CP$-even Higgs  bosons,  one $\CP$-odd  Higgs
boson,  and two  charged  Higgs  bosons.  The  NMSSM  Higgs sector  is
extended  by  an  additional   complex  superfield  leading  to  three
$\CP$-even,  two  $\CP$-odd  and  two  charged  Higgs  bosons  in  the
$\CP$-conserving case.   Contrary to the  SM, the masses of  the Higgs
bosons  can be  predicted in  terms of  the parameters  of the  model.
While supersymmetric  relations lead  to an upper  mass bound  for the
light  $\CP$-even  Higgs  boson  below  125~GeV  at  tree  level,  the
inclusion  of  higher-order corrections  can  shift  the mass  to  the
observed  value.   However,  both  in  the MSSM  and  the  NMSSM,  the
theoretical uncertainties of  the current predictions for  the mass of
the SM-like Higgs boson are significantly larger than the experimental
error. For  the MSSM detailed estimates  for theoretical uncertainties
of     the    Higgs-mass     predictions     are    available,     see
e.g.~\cite{Degrassi:2002fi, Allanach:2004rh}.   Automated estimates of
the theoretical  uncertainties depending  on the  considered parameter
point    within   the    MSSM   can,    e.g.,   be    performed   with
\FH~\cite{Heinemeyer:1998yj,    Heinemeyer:1998np,    Degrassi:2002fi,
  Frank:2006yh,      Hahn:2009zz,     Hahn:2013ria,      Bahl:2016brp,
  feynhiggs-www}.

For the  NMSSM several public  spectrum generators are  available that
provide   an  automated   calculation  of   the  Higgs-boson   masses:
\texttt{FlexibleSUSY}~\cite{Athron:2014yba},
\texttt{FlexibleEFTHiggs}~\cite{Athron:2016fuq},
\texttt{NMSSMCALC}~\cite{Baglio:2013iia,                King:2015oxa},
\texttt{NMSSMTools}~\cite{Ellwanger:2004xm,          Ellwanger:2005dv,
  Ellwanger:2006rn},          \texttt{SOFTSUSY}~\cite{Allanach:2001kg,
  Allanach:2013kza,                Allanach:2014nba}               and
\texttt{SPheno}~\cite{Porod:2003um,   Porod:2011nf}.     The   results
obtained by the  different codes for the same set  of input parameters
can differ by several $\giga\electronvolt$~\cite{Staub:2015aea}.

A first  step towards investigating the  theoretical uncertainties for
NMSSM  Higgs-mass predictions  focusing on  calculations using  a pure
$\DRbar$        renormalization        has       been        performed
in~\cite{Staub:2015aea}. In this  publication the aforementioned tools
(except for \texttt{FlexibleEFTHiggs}, which  did not exist then) were
used to  calculate NMSSM Higgs-boson  masses for six  sample scenarios
with different  physical properties.  The sources  for the differences
between the codes have been  identified, and after modifying the codes
to  use   the  same  approximations   they  agree  at  the   level  of
\order{10~\mev}  (for  the  same  set  of  higher-order  corrections).
However,  this  technical  agreement  does   not  allow  one  to  draw
conclusions on  the remaining  theoretical uncertainties  from unknown
higher-order    corrections.     In   particular,    the    comparison
in~\cite{Staub:2015aea} did not account for differences resulting from
the use of  different renormalization schemes. Among  the tested tools
only   \texttt{NMSSMCALC}   offers   the   option   to   use   another
renormalization scheme, namely a  mixed $\DRbar$/on-shell (OS) scheme.
The  results in  this scheme  were  not considered  in the  comparison
of~\cite{Staub:2015aea}.  In  the present  work, we will  address this
issue by  comparing codes  incorporating a $\DRbar$/OS  scheme, namely
\texttt{NMSSMCALC}     and    the     NMSSM-extended    version     of
\texttt{FeynHiggs} \cite{Drechsel:2016jdg}.  We stick here exclusively
to  the  codes  \texttt{NMSSMCALC} and  \texttt{FeynHiggs},  with  the
latter  applying  a  mixed $\DRbar$/OS  renormalization  scheme  only.
Concerning the comparison between the $\DRbar$/OS mixed scheme and the
pure   $\DRbar$  scheme,   we  investigate   the  differences   within
\texttt{NMSSMCALC} for the two  renormalization schemes.  The analysis
performed in this  paper yields important ingredients  for an estimate
of the  remaining theoretical uncertainties from  unknown higher-order
corrections for the Higgs-boson mass calculations in the NMSSM.

The paper is organised as  follows.  In \refse{sec:NMSSM} we introduce
our notation for the  relevant NMSSM parameters.  In \refse{sec:Codes}
we describe  the codes  \NC\ and \NFH\  together with  the differences
between them. The analyzed numerical  scenarios and their treatment is
described  in \refse{sec:Scenarios}.   The  obtained  results for  the
masses and  mixing matrices are  discussed, and their  differences are
analyzed  in \refse{sec:Results}.   The  conclusions can  be found  in
\refse{sec:conclusion}.

%% file: sections/NMSSM.tex
\section{The relevant NMSSM sectors }
\label{sec:NMSSM}
The   superpotential   of  the   NMSSM   for   the  third   generation
fermions/sfermions reads
\begin{align}
\label{eq:superpot}
  W = 
 Y_{b} \left(\hat{H}_1\cdot \hat{Q}_3 \right)\hat{d}_3 +Y_\tau \left(\hat{H}_1\cdot \hat{L}_3 \right)\hat{e}_3
 - \yt \left(\hat{H}_2\cdot \hat{Q}_3 \right)\hat{u}_3 
  + \lambda \hat{S} \left(\hat{H}_2 \cdot \hat{H}_1 \right) + \frac{1}{3} \kappa \hat{S}^3,
\end{align}
\noindent
with  the  left-handed  quark  and  lepton  superfields,  $\hat{Q}_3$,
$\hat{L}_3$, and the right-handed  ones, $\hat{u}_3$, $\hat{d}_3$, and
$\hat{e}_3$,  exemplary  for  all  three generations,  and  the  Higgs
superfields    $\hat{H}_1$,    $\hat{H}_2$   and    $\hat{S}$.     The
$\SUL$-invariant product is  denoted by a dot. Since we  will focus on
the  \cp-conserving  NMSSM in  this  comparison,  all the  Yukawa-type
couplings  $\yt$, $Y_{b}$,  $Y_\tau$,  $\lambda$ and  $\kappa$ can  be
chosen as real parameters. The  scalar components $H_1$, $H_2$ and $S$
of the  Higgs doublet and  singlet superfields can be  decomposed into
\cp-even   and  \cp-odd   neutral   scalars   $\phi_x$  and   $\chi_x$
($x=1,2,s$), respectively, and  charged states $\phi^\pm_i$ ($i=1,2$).
After expansion about their  vacuum expectation values (vevs) $\langle
H_i \rangle$ ($i=1,2$) and $\langle S \rangle$, they read
\begin{align}
 H_1 = \begin{pmatrix} \langle H_1\rangle +\frac{1}{\sqrt{2}}\left(
  \phi_1 \pm i\chi_1\right) \\ 
  \pm \phi_1^- \end{pmatrix} , \quad
 H_2 = \begin{pmatrix} \phi_2^+ \\ \langle H_2 \rangle + \frac{1}{\sqrt{2}}\left(\phi_2+i\chi_2\right) \end{pmatrix} , \quad
 S = \langle S \rangle +\frac{1}{\sqrt{2}}\left( \phi_s+i\chi_s\right) .
\end{align}
\noindent
The plus  sign in the doublet  $H_1$ refers to the  convention used in
\NC, the minus  sign to the one used in  \NFH. Due to \cp-conservation
the  vevs are  real.  Since  $\hat{S}$ transforms  as  a singlet,  the
$D$-terms remain identical to the ones from the MSSM.  Compared to the
\cp-conserving  MSSM the  superpotential of  the \cp-conserving  NMSSM
contains additional  dimensionless parameters $\lambda$  and $\kappa$,
while the $\mu$-term is absent. This term is generated effectively via
the vev of the singlet field,
\begin{align}
 \mueff = \lambda \left<S\right> .
\end{align}
\noindent
It should be noted that there  are two common conventions for defining
the  vacuum expectation  values, $\left<S\right>  = v_s/\sqrt{2}$  and
$\left<H_i\right>  =  v_i/\sqrt{2}$,  or $\left<S\right>  =  v_s$  and
$\left<H_i\right> = v_i$, respectively.  Both  are allowed by the SLHA
conventions~\cite{Allanach:2008qq,    Skands:2003cj}.    The    latter
convention is used by \FH, while the  former is used by \NC. As in the
MSSM it is convenient to define the ratio
\begin{align}
 \tb = \frac{v_2}{v_1} .
\end{align}
\noindent
Soft  SUSY-breaking in  the  NMSSM  gives rise  to  the  real (in  the
\cp-conserving   case)   trilinear   soft   SUSY-breaking   parameters
$A_\lambda$ and $A_\kappa$, as well  as to the soft-SUSY breaking mass
term  $m_S^2$ of  the scalar  singlet field.   Together with  the soft
SUSY-breaking Lagrangian of the MSSM we have
\begin{align}
 \mathcal{L}_{\text{soft}} = 
 \mathcal{L}_{\text{soft,MSSM}}
 - m_S^2 |S|^2 - \left[
  \lambda A_\lambda S \left(H_2 \cdot H_1\right) + \frac{1}{3} \kappa A_\kappa S^3
  + \text{h.c.}
  \right].
\end{align}
\noindent
The  MSSM  soft  SUSY-breaking  Lagrangian, exemplary  for  the  third
generation, reads
\begin{align}
  {\cal L}_{\text{soft,MSSM}} =
  &- m_1^2 H_{1}^\dagger H_{1} - m_2^2 H_{2}^\dagger H_{2} 
  - M_{\tilde{Q}_3}^2 \tilde{Q}_3^\dagger \tilde{Q}_3 
  - M_{\tilde{L}_3}^2 \tilde{L}_3^\dagger \tilde{L}_3 
  \nonumber \\
  &- M_{\tilde{t}_R}^2 \tilde{u}_{3}^* \tilde{u}_{3} 
  - M_{\tilde{b}_R}^2 \tilde{d}_{3}^* \tilde{d}_{3} 
  - M_{\tilde{\tau}_R}^2 \tilde{e}_{3}^* \tilde{e}_{3} 
  \nonumber\\
  &-([
    Y_\tau A_\tau (H_1 \cdot \tilde{L}_3) \tilde{e}_3^* 
    + Y_{b} A_{b} (H_1 \cdot\tilde{Q}_3) \tilde{d}_{3}^* 
    - Y_{t} A_{t} (H_2 \cdot \tilde{Q}_3) \tilde{u}^*_3
  ] + h.c.)
  \nonumber\\
  &- \frac{1}{2} (
  M_1 \tilde{B} \tilde{B} 
  + M_2 \tilde{W}_i \tilde{W}_i
  + M_3 \tilde{G} \tilde{G} + h.c.)
  \;,
\end{align}
\noindent
where the tilde in the first  three lines denotes the scalar component
of     the     corresponding     superfield,     $m_1^2$,     $m_2^2$,
$M_{\tilde{Q}_3}^2$,     $M_{\tilde{t}_R}^2$,     $M_{\tilde{b}_R}^2$,
$M_{\tilde{L}_3}^2$  and  $M_{\tilde{\tau}_R}^2$   are  the  soft-SUSY
breaking mass  parameters for  the Higgs bosons,  the squarks  and the
sleptons,  respectively,  and  $A_t$,   $A_b$  and  $A_\tau$  are  the
soft-SUSY breaking  trilinear couplings  of the squarks  and sleptons.
The last line summarises the soft SUSY-breaking gaugino mass terms for
the  \UY, \SUL,  and \SUc\  gaugino fields  $\tilde{B}$, $\tilde{W}_i$
($i=1,2,3$) and  $\tilde{G}$ with the gaugino  mass parameters, $M_1$,
$M_2$ and $M_3$. The Higgs potential $V_H$ can be written in powers of
the fields,
\begin{align}
  \label{eq:HiggsPot}
  &V_H =
  \ldots - T_{\phi_1} \phi_1 - T_{\phi_2} \phi_2 - T_{\phi_S} \phi_s \\
  &\quad + \frac{1}{2} \begin{pmatrix} \phi_1,\phi_2, \phi_s \end{pmatrix}
  \mathbf{M}_{\phi\phi} \begin{pmatrix} \phi_1 \\ \phi_2 \\ \phi_s \end{pmatrix} +
  \frac{1}{2} \begin{pmatrix} \chi_1, \chi_2, \chi_s \end{pmatrix}
  \mathbf{M}_{\chi\chi}
  \begin{pmatrix} \chi_1 \\ \chi_2 \\ \chi_s \end{pmatrix} +
  \begin{pmatrix} \phi^-_1,\phi^-_2 \end{pmatrix}
  \mathbf{M}_{\phi^\pm\phi^\pm}
  \begin{pmatrix} \phi^+_1 \\ \phi^+_2 \end{pmatrix} + \ldots,
  \nonumber
\end{align}
\noindent
where  the coefficients  linear and  bilinear  in the  fields are  the
tadpole  parameters $T_{\phi_1}$,  $T_{\phi_2}$, $T_{\phi_S}$  and the
mass              matrices             $\text{\textbf{M}}_{\phi\phi}$,
$\text{\textbf{M}}_{\chi\chi}$                                     and
$\text{\textbf{M}}_{\phi^\pm\phi^\pm}$, respectively.  The dots denote
constant terms and terms trilinear and quartic in the fields.

For   the  \cp-conserving   case  the   mixing  into   the  mass   and
\cp\ eigenstates  can be  described at lowest  order by  the following
unitary transformations
\begin{align}
 \label{eq:MixMatrices}
 \begin{pmatrix} h_1 \\ h_2 \\ h_3 \end{pmatrix}
 = \mathbf{U}_{e(0)}
 \begin{pmatrix} \phi_1 \\ \phi_2 \\ \phi_s \end{pmatrix}, \quad
 \begin{pmatrix} A_1 \\ A_2 \\ G^0 \end{pmatrix}
 = \mathbf{U}_{o(0)}
 \begin{pmatrix} \chi_1 \\ \chi_2 \\ \chi_s \end{pmatrix}, \quad
 \begin{pmatrix} H^\pm \\ G^\pm \end{pmatrix}
 = \mathbf{U}_{c(0)}
 \begin{pmatrix} \phi^\pm_1 \\ \phi^\pm_2 \end{pmatrix} .
\end{align}
\noindent
The new fields  correspond to the five neutral Higgs  bosons $h_i$ and
$A_j$, the charged Higgs pair  $H^\pm$, and the Goldstone bosons $G^0$
and  $G^\pm$. The  matrices $\mathbf{U}_{\{e,o,c\}(0)}$  transform the
Higgs  fields   such  that  the  mass   matrices  $  \mathbf{M}_{hh}$,
$\mathbf{M}_{AA}$, and $\mathbf{M}_{H^\pm  H^\mp}$ are diagonalized at
tree level,
\begin{align}
 \mathbf{M}_{hh} =
 \mathbf{U}_{e(0)} \mathbf{M}_{\phi\phi} \mathbf{U}_{e(0)}^\dagger
 ,\
 \mathbf{M}_{AA} = \mathbf{U}_{o(0)} \mathbf{M}_{\chi\chi} \mathbf{U}_{o(0)}^\dagger
 ,\
 \mathbf{M}_{H^\pm H^\mp} = \mathbf{U}_{c(0)} \mathbf{M}_{\phi^\pm\phi^\pm} \mathbf{U}_{c(0)}^\dagger.
\end{align}

%% file: sections/codes.tex
\section{The two codes: \NC\ and \FH}
\label{sec:Codes}

In this section  we will give a brief overview  about the higher-order
corrections to Higgs-boson  masses included in \FH\  and \NC, together
with the different renormalization schemes employed.  We will restrict
ourselves here and in the following to the CP-even Higgs sector.


\subsection{Incorporation of higher-order contributions}
\label{sec:calcho}

The masses of the \cp-even Higgs  bosons are obtained from the complex
poles of the full propagator matrix. The inverse propagator matrix for
the three \cp-even Higgs  bosons $h_i$ from eq.~\eqref{eq:MixMatrices}
is a $3 \times 3$ matrix which reads
\begin{align}
 \Delta^{-1}{\left(k^2\right)} = \mathi
 \left[k^2\mathds{1} - \mathbf{M}_{hh}
 + \mathbf{\hat{\Sigma}}_{hh}{\left(k^2\right)}
 \right].
\end{align}
\noindent
Here,  $\hat{\Sigma}_{hh}$  denotes  the matrix  of  the  renormalized
self-energy corrections of the CP-even Higgs fields. The three complex
poles of the propagator in the  \cp-even Higgs sector are given by the
values of the external momentum $k^2$ for which the determinant of the
inverse propagator-matrix vanishes,
\begin{align}
 \det\left[
 \Delta^{-1}{\left(k^2\right)}
 \right]_{k^2 = M_{h_i}^2 - \mathi \Gamma_{h_i} M_{h_i}} \overset{!}{=} 0
 , \ i \in \{1,2,3\}.
\end{align}
\noindent
The real  parts of the three  poles are identified with  the square of
the Higgs-boson  masses $M_{h_i}$  in the  \cp-even sector,  while the
imaginary  parts  include  their  total  widths  $\Gamma_{h_i}$.   The
renormalized   self-energy   matrix  $\mathbf{\hat{\Sigma}}_{hh}$   at
one-loop  order is  evaluated in  \NC~\cite{Ender:2011qh, Graf:2012hh}
and  \NFH~\cite{Drechsel:2016jdg}  by  taking into  account  the  full
contributions  from  the  NMSSM (differences  in  the  renormalization
schemes are  discussed below).   At two-loop  order \texttt{NMSSMCALC}
includes the  leading (S)QCD corrections  from the top/stop  sector of
\order{\alt\als}   in   the  NMSSM~\cite{Muhlleitner:2014vsa},   while
\NFH\ uses all available corrections  from the MSSM, that are included
in  the  MSSM-version   of  \FH~\cite{Heinemeyer:1998yj,  Hahn:2009zz,
  Heinemeyer:1998np,   Degrassi:2002fi,  Frank:2006yh,   Hahn:2013ria,
  Bahl:2016brp, feynhiggs-www}, as an approximation%
\footnote{Updates beyond the \FH\ version 2.10.2 (used for
this comparison) also take into account momentum dependent two-loop 
contributions~\cite{Borowka:2014wla, Degrassi:2014pfa} 
and improved resummations of large logarithmic
corrections~\cite{Bahl:2016brp}. These updates are not relevant for the
comparison between the two codes up to \order{\alt\als}.}%
,
\begin{subequations}
 \label{eq:SEapprox}
 \begin{align}
 \NC:\
 &\mathbf{\hat{\Sigma}}_{hh}^{\tt NC}{\left(k^2\right)}
 =
 \left.
 \mathbf{\hat{\Sigma}}^{(\text{1L})}_{hh}{\left(k^2\right)}
 \right|^{\text{NMSSM}} +
 \left.
 \mathbf{\hat{\Sigma}}^{(\text{2L})}_{hh}{\left(k^2\right)}
 \right|_{k^2=0}^{\text{NMSSM }\order{\alt\als}},
 \\
 \FH:\
 &\mathbf{\hat{\Sigma}}_{hh}^{\FH}{\left(k^2\right)}
 =
 \left.
 \mathbf{\hat{\Sigma}}^{(\text{1L})}_{hh}{\left(k^2\right)}
 \right|^{\text{NMSSM}} +
 \Big[
 \left.\mathbf{\hat{\Sigma}}^{(\text{2L})}_{hh}{\left(k^2\right)}
 \right|^{\order{\alt\als}}+
 \order{\alt^2,\alb\als,\alt\alb}
 \nonumber\\
 & \mbox{}\hspace{80mm} + \mbox{resummed logs} \Big]_{k^2=0}^{\text{MSSM}}.
 \end{align}
\end{subequations}
\noindent
In  order to  facilitate the  comparison  between \NC\  and \NFH\,  at
two-loop   order   we   only   include   the   MSSM   corrections   of
\order{\alt\als} in \NFH, if not stated otherwise.

The  mixing matrix  elements  including  higher-order corrections  are
denoted  by \ZH{ij}.  Here and in the following we will suppress the loop order of the mixing
  matrix, but specify it in the text. They are  given  by the  unitary matrices  that
diagonalize the  mass-matrix at  tree level,  $\mathbf{U}_{e(0)}$, and
the   loop-corrected  mass   matrix   for   zero  external   momentum,
$\mathbf{U}_{e(i)}$ with $i$ denoting the loop order,
\begin{align}
  \mathbf{U^{h}} = \mathbf{U}_{e(i)} \mathbf{U}_{e(0)},
  \label{Uh}
\end{align}
\noindent
where
\begin{align}
  {\rm diag}{\left(M^2_{h_1,0}, M^2_{h_2,0}, M^2_{h_3,0}\right)}
  =
  \mathbf{U}_{e(i)}
  \left[{\rm diag}{\left(m^2_{h_1}, m^2_{h_2}, m^2_{h_3}\right)} +
    \mathbf{\hat{\Sigma}}_{hh}^{(i)} {\left(k^2=0\right)}\right]
  \mathbf{U}_{e(i)}^\dagger.
\end{align}
Here,  $m_{h_j}$  and  $M_{h_j,0}$  with   $j  =  1,2,3,$  denote  the
Higgs-boson masses  at tree-level  and at higher  order, respectively,
i.e.~including up to one-loop corrections for $i=1$ and up to two-loop
contributions  for $i=2$,  with vanishing  external momentum  $k^2=0$.
The  evaluation  of the  mixing  matrices  at zero  external  momentum
ensures the  unitarity of  the mixing  matrices.  The  mixing matrices
considered here  differ from  the wave function  normalization factors
for  external Higgs  bosons in  an  S-matrix element.  The latter  are
evaluated  at  the  complex  poles  of  the  propagators  and  form  a
non-unitary  matrix. We  found that  the differences  between the  two
types of matrices are small for most of the scenarios.


\subsection{Renormalization scheme: Higgs- and electroweak sectors}
\label{sec:RenSchemeHEW}

The independent parameters appearing in  the linear and bilinear terms
of the Higgs potential in  \refeq{eq:HiggsPot} have to be renormalized
for  the evaluation  of  higher-order corrections  to the  Higgs-boson
masses.   \NC\  and \NFH\  offer  different  choices  for the  set  of
independent    parameters    and     the    applied    renormalization
schemes~\cite{Drechsel:2016jdg,       Ender:2011qh,       Graf:2012hh,
  Muhlleitner:2014vsa}.   For the  presented  work the  set of  common
independent parameters in  the Higgs sector for the  comparison of the
mixed \DRbar/OS renormalization schemes in the two codes reads
\begin{align}
 \underbrace{\MZ,\ \MW,\ \MHp, T_{\phi_{\{1,2,s\}}}}_{\text{on-shell}}, 
\underbrace{\TB,\ \lambda,\ \mueff,\ \kappa,\ \Ak}_{\DRbar},
\end{align}
\noindent
with the  applied renormalization  scheme. In \NC\  the scheme  can be
varied: If $\MHp$  is set as input parameter, it  will be renormalized
OS,   if   instead   the  trilinear   soft-SUSY   breaking   parameter
$A_{\lambda}$ is used as input  parameter it is renormalized $\DRbar$,
see  \refse{sec:DRbar}.  The  gauge-boson masses,  however, are  still
renormalized OS,  and the  tadpole coefficients are  renormalized such
that the renormalized tadpoles vanish.

Using the former option, $\MHp$ as input, up to the one-loop level the
two codes differ only by their treatment of the renormalization of the
coupling constant  $\alpha$ in the electromagnetic  sector.  While for
\NC\  $\alpha$  is  renormalized  to $\alpha(\MZ)$,  \NFH\  employs  a
dependent renormalization scheme  (employing a \DRbar\ renormalization
of $v$) with a subsequent reparametrization~\cite{Drechsel:2016jdg} to
the value  $\alpha_{G_F}$, derived from  the Fermi constant  $G_F$ (to
match  exactly the  \FH\ MSSM  evaluation  in the  MSSM limit).   This
difference  in the  treatment  of the  charge  renormalization at  the
one-loop level  is formally an  effect of electroweak  two-loop order.
In  the Higgs-boson  mass  calculation in  the  MSSM-limit the  charge
renormalization  constant drops  out at  the discussed  levels of  the
calculation  and thus  its  impact  is a  genuine  NMSSM effect.   The
differences  between  \NC\  and  \NFH\ in  the  contributions  to  the
Higgs-boson self-energies at the one-loop  and the two-loop level (see
below) are summarized in \refta{tab:DiffNcNfhVanilla}.

\begin{table}[h]
 \centering
 \begin{tabular}{lcccc}
 & \NC & & \NFH
 \\\toprule
 \multirow{2}{*}{1-loop} & 
 $\alpha{\left(M_Z\right)}$ & \multirow{2}{*}{$\leftrightarrow$} & $\alpha_{G_F}$
\\
 & renormalized & & reparametrized
 \\
 \midrule
 \multirow{2}{*}{2-loop} & 
 $\als^{\DRbar}\left(Q_{\rm input}\right)$ & 
 $\leftrightarrow$ &
 $\als^{\MSbar}\left(\mt\right)$ 
 \\
 &
 \multirow{2}{*}{NMSSM \order{\alt\als}} &
 \multirow{2}{*}{$\leftrightarrow$} &
 MSSM \order{\alt\als, \alt^2, \alb\als, \alt\alb}
 \\
 & & &
 + resummed logarithms
 \\\bottomrule
 \end{tabular}
 \caption{Calculational differences  between the original  versions of
   \NC\ and \NFH\ as used for this comparison.  The values applied for
   the electromagnetic and strong coupling constants are stated, where
   $\alpha_{G_F}$  denotes   the  electromagnetic   coupling  constant
   calculated from the Fermi constant.}
\label{tab:DiffNcNfhVanilla}
\end{table}


\subsection{Renormalization scheme: top/stop sector}
\label{sec:RenSchemeTST}

For the two-loop  \order{\alt\als} corrections the top  quark mass and
the stop parameters need to be renormalized.  In \NC\ either the OS or
the  \DRbar\ renormalization  scheme for  the top/stop  sector can  be
used. Apart from section~\ref{sec:DRbar}, where we indicate explicitly
the renormalization  scheme of the  top/stop sector, we employ  the OS
scheme in \NC\  throughout this work.  In \NFH\ the  OS scheme is used
throughout for  the parameters in  the stop sector. For  the top-quark
mass,  either the  OS or  the  \DRbar\ renormalization  scheme can  be
chosen in \NFH, and a further  option is to use a reparametrization of
the OS result in terms of the \MSbar\ mass of SM QCD.

In  both programs,  the  OS  scheme is  defined  by applying  on-shell
conditions for the  respective masses, i.e.\ the  top-quark mass $m_t$
and the top squark  masses $m_{\tilde{t}_1}$ and $m_{\tilde{t}_2}$.  A
fourth renormalization condition  fixes the mixing of  the squarks and
can be  identified with a  condition for  the stop mixing  angle.  The
resulting counterterms have the same form  as in the MSSM, and details
can be  found in refs.~\cite{Heinemeyer:2007aq,  Degrassi:2001yf}.  No
additional counterterms  of the  sbottom sector  are needed  since the
bottom mass is set to zero in the charged Higgs self-energies.

\NC\   uses  the   soft-SUSY  breaking   masses  of   left-handed  and
right-handed fields,  $\msqd$ and $M_{\tilde{t}_{R}}$, as  well as the
trilinear  coupling  $A_t$  and   calculates  counterterms  for  these
parameters  corresponding  to   the  above  mentioned  renormalization
conditions. In this  way, switching from OS to  \DRbar\ parameters can
easily be  done, see ref.~\cite{Muhlleitner:2014vsa}.  \NFH\  uses the
same numerical input  values of the soft-SUSY  breaking parameters and
the same OS  conditions. Counterterms are employed in  the stop sector
for the stop masses and the stop mixing angle.


\subsection{Treatment of QCD corrections}

In   \NC\  the   $\DRbar$-value  of   the  strong   coupling  constant
$\als^{\DRbar}$ is calculated at the input scale $Q$ of the parameters
specified  in the  SLHA  input  file by  applying  the formulae  given
in~\cite{Chetyrkin:2000yt,Baer:2002ek}.   In \NFH\  the $\MSbar$-value
of the strong  coupling constant $\als^{\MSbar}$ is  calculated at the
scale $\mt$.   In both codes  the obtained value is  subsequently used
for the  evaluation of the  two-loop contributions to the  Higgs boson
masses.   As  stressed  above,  \NC\ includes  corrections  of  up  to
\order{\alt\als}, and consequently for  our comparison we restrict the
\NFH\ evaluation to this order as well.  The treatment of the two-loop
contributions    is     summarized    in    the    lower     row    of
\refta{tab:DiffNcNfhVanilla}.

%% file: sections/scenarios.tex
\section{Description of the scenarios}
\label{sec:Scenarios}

\subsection{The five test-point scenarios}

In \citere{Staub:2015aea}, six test-point (TP) scenarios were proposed
for the comparison of the Higgs-mass predictions obtained by different
tools using the  \DRbar\ scheme.  They will also be  employed here for
the  comparison  between  \NC\  and   \NFH\,  to  facilitate  a  later
comparison with  ref.~\cite{Staub:2015aea}. The definitions of  the TP
scenarios  are recapitulated  in  \refta{tab:TPDefinition}, where  all
parameters are  given at the indicated  scale, both at the  high scale
\MS\, at which  they were originally defined, and at  the scale of the
top-quark on-shell mass $\MTO$.%
\footnote{In   the   original   definition   of   the   scenarios   in
  ref.~\cite{Staub:2015aea} $\TB$ is given  by its \DRbar-value at the
  scale of the $Z$-boson mass $\MZ$, while in \refta{tab:TPDefinition}
  we  give  the corresponding  \DRbar-value  at  the indicated  scale,
  either \MS\  or \MTO. The values  for $\TB$ given here  are obtained
  with  \texttt{FlexibleSUSY}   as  described   in  the   text.}   For
completeness  we  repeat  the  different physical  features  of  these
scenarios as given in \citere{Staub:2015aea}:
\begin{itemize}
\item[TP1:] MSSM-like point.
\item[TP2:] MSSM-like point with large stop splitting.
\item[TP3:]  Point  with light  singlet  and  $\lambda$ close  to  the
  perturbativity limit.
\item[TP4:]  Point  with heavy  singlet  and  $\lambda$ close  to  the
  perturbativity limit.
\item[TP5:]  Point with  slightly lighter  singlet. Additional  matter
  needed for perturbativity; inspired by~\cite{King:2012is}.
\end{itemize}
\noindent
The scenario TP6 of \citere{Staub:2015aea}  is characterized by a very
large value  of $\lambda$.  It  will be omitted from  this comparison,
since  the corrections  beyond the  $\mathcal{O}{(\alpha_t \alpha_s)}$
approximation        can       be        sizeable       in        this
case~\cite{Goodsell:2014pla,Staub:2015aea}.      Furthermore,     this
scenario requires  new physics well below  the GUT scale to  avoid the
non-perturbative    regime.     All     TP    scenarios,    using    a
\DRbar\  renormalization, contain  a SM-like  Higgs-field with  a mass
predicted at the two-loop level of around $125~\gev$.

\begin{table}
  \scriptsize
  \centering
  \begin{tabular}{crrrrrrrrrrrrrr}
    & $Q$ & \TB & $\lambda$ & $\kappa$ & $\Al$ & $\Ak$ & $\mueff$ &
    $M_1$ & $M_2$ & $M_3$ & $A_t$ & $A_b$ & $M_{\tilde{Q}_3}$ & $M_{\tilde{t}_R}$
    \\\toprule
    \multirow{2}[4]{*}{TP1} &
    \MS & $9.599$ & $0.100$ & $0.100$ & $-10.00$ & $-10.00$ & $900.0$ & $500.0$ & 
    $1000$ & $3000$ & $3000$ & $0$ & $1500$ & $1500$
    \\\cmidrule{2-15}
    & $\mt$ & $9.903$ & $0.098$ & $0.100$ & $-198.6$ & $-9.738$ & $886.6$ & $478.1$ & 
    $979.4$ & $3261$ & $2154$ & $-585.2$ & $1907$ & $1838$
    \\\midrule\midrule
    \multirow{2}[4]{*}{TP2} &
    \MS & $9.621$ & $0.050$ & $0.100$ & $-200.0$ & $-200.0$ & $1500$ & 
    $1000$ & $2000$ & $2500$ & $-2900$ & $0$ & $2500$ & $500$
    \\\cmidrule{2-15}
    & $\mt$ & --- & --- & --- & --- & --- & --- & --- & --- & --- & --- & --- & --- & ---
    \\\midrule\midrule
    \multirow{2}[4]{*}{TP3} &
    \MS & $2.881$ & $0.670$ & $0.100$ & $650.0$ & $-10.00$ & $200.0$ &
    $200.0$ & $400.0$ & $2000$ & $1000$ & $1000$ & $1000$ & $1000$
    \\\cmidrule{2-15}
    & $\mt$ & $2.967$ & $0.648$ & $0.097$ & $574.0$ & $-43.68$ & $195.3$ &
    $192.4$ & $391.9$ & $2145$ & $619.8$ & $691.5$ & $1233$ & $1211$
    \\\midrule\midrule
    \multirow{2}[4]{*}{TP4} &
    \MS & $1.920$ & $0.670$ & $0.200$ & $405.0$ & $0$ & $200.0$ & $120.0$ & $200.0$ &
    $1500$ & $1000$ & $1000$ & $750.0$ & $750.0$
    \\\cmidrule{2-15}
    & $\mt$ & $1.975$ & $0.649$ & $0.195$ & $344.3$ & $0.195$ & $195.3$ & $116.0$
    & $195.7$ & $1591$ & $726.5$ & $801.5$ & $891.9$ & $874.4$
    \\\midrule\midrule
    \multirow{2}[4]{*}{TP5} &
    \MS & $2.864$ & $0.670$ & $0.200$ & $570.0$ & $-25.00$ & $200.0$ &
    $135.0$ & $200.0$ & $1400$ & $0$ & $0$ & $1500$ & $1500$
    \\\cmidrule{2-15}
    & $\mt$ & $2.967$ & $0.643$ & $0.193$ & $549.3$ & $-63.47$ & $194.4$ & 
    $128.7$ & $194.3$ & $1526$ & $-234.8$ & $-251.9$ & $1579$ & $1546$
    \\\bottomrule
  \end{tabular}
  \caption{Definition of  the TP scenarios.  All  parameters are given
    as   \DRbar\  parameters   at  the   indicated  scale   $Q$.   All
    dimensionful  parameters  are  given   in  $\gev$.  The  remaining
    parameters,  common  to all  points,  are  the soft  SUSY-breaking
    parameters  in  the  sfermion mass  matrices,  $M_{\tilde{L}_i}  =
    M_{\tilde{Q}_j}  =  M_{\tilde{f}_R}=M_{\tilde{b}_R} =  1500~\gev$,
    where $i =1,2,3$,  $j=1,2$, $f=e, \mu, \tau, d, s,  u, c$, and the
    trilinear sfermion-Higgs coupling $A_f = 0$~GeV at the high scale,
    $\MS  = \tfrac{1}{2}(M_{\tilde  Q_3} +  M_{\tilde t_R})$.   At the
    scale  $Q=\mt=172.9~\gev$ we  use the  corresponding SUSY-breaking
    parameters but  evolved to  the scale  $Q=\mt$.  The  scenario TP2
    yields tachyonic stop-masses at the scale \MTO.}
  \label{tab:TPDefinition}
\end{table}

The measured value of the discovered Higgs boson of $\sim 125~\gev$ is
at the  weak scale.  The  diagrammatic corrections to  the Higgs-boson
self-energies  in \NC\  and  \NFH\  are such  that  the full  particle
spectrum of the model is  incorporated in the loop contributions. This
approach is motivated by scenarios  where the SUSY scale is relatively
low, i.e.\ not  widely separated from the weak scale,  and where there
is no  large hierarchy  among the SUSY  particle masses.   Within this
context,  the   diagrammatic  approach  yields  general   results  for
arbitrary values  of the  involved parameters. In  contrast, effective
field theory  (EFT) methods  are designed for  the treatment  of large
scale splittings within the calculation.%
\footnote{As stated above, the  contributions from resummed logarithms
  in  \FH\  obtained  using  EFT  methods  are  not  included  in  our
  comparison.}  Within  the context of the  diagrammatic calculations,
we  investigate two  possible  treatments  of the  scale  of the  SUSY
parameters.   We will  perform the  conversion  of the  \DRbar\ to  OS
parameters  and  (for  the  calculation in  the  \DRbar\  scheme)  the
evaluation of the  Higgs-boson masses for the given  scenarios both at
their original scale $\MS\ =  \tfrac{1}{2} (M_{\tilde Q_3} + M_{\tilde
  t_R})$, the arithmetic  mean of the two  diagonal soft SUSY-breaking
mass parameters in the scalar top mass matrix, as well as at the scale
of the OS top-quark mass $\MTO = 172.9~\gev$.  In the latter case this
means in particular  that the parameters are first  evolved from their
original scale \MS\ to the  scale \MTO\ with {\tt FlexibleSUSY} before
they are converted to OS parameters.

The    parameters    at    the    scale    \MTO\    are    given    in
\refta{tab:TPDefinition}, together with the original parameters at the
scale \MS. The scenario TP2  yields tachyonic stop-masses at the scale
\MTO. Consequently, TP2 is evaluated only at the scale~\MS.


\subsection{Conversion from \boldmath{\DRbar} to OS parameters}
\label{sec:drbaros}

In \NC\ it  is possible to perform calculations with  either the OS or
the  \DRbar\  renormalization  scheme  in  the  top/stop  sector.   In
\NFH\ the OS scheme is mandatory  for the scalar top quarks, while for
the  top-quark  mass   the  OS  scheme,  the  \DRbar\   scheme  and  a
reparametrization of the OS result in  terms of the \MSbar\ mass of SM
QCD can be chosen.  Both codes are capable of converting \DRbar\ input
parameters  at their  given  scale  into OS  parameters  by using  the
well-known OS shifts of the MSSM~\cite{Carena:2000dp,Degrassi:2001yf}%
\footnote{Since  only  corrections  of   up  to  \order{\alt\als}  are
  discussed in this work, only one-loop shifts of \order{\als} for the
  scalar top  sector are  necessary. These are  identical in the
  MSSM and the NMSSM.}.   For our comparison we used the  routines of only
one  code,  \NC, in  order  to  prevent  effects that  originate  from
different implementations  of these  shifts.  The  shifts in  \NC\ are
computed as
\begin{equation}
  X^{(\OS)}=X^{(\DRbar)}-\delta X^{\rm fin}
  \hspace*{1cm}\text{  with   } \hspace*{1cm}
  X= \msqd, \, M_{\tilde{t}_{R}}, \, \At \,,
\end{equation}
\noindent
where  $X^{(\DRbar)}$ and  $\delta  X^{\rm fin}$  both  depend on  the
\DRbar\ scale. The shift $\delta  X^{\rm fin}$ denotes the finite part
of the  respective counterterm obtained in  the OS scheme as  given in
ref.~\cite{Muhlleitner:2014vsa}.   They  are computed  iteratively  by
inserting the obtained OS parameters, until convergence is reached.

\begin{table}
  \scriptsize
  \centering
  \begin{tabular}{crrrrrrrrrr}
    & \multicolumn{2}{c}{TP1} & \multicolumn{2}{c}{TP2} & \multicolumn{2}{c}{TP3}
    & \multicolumn{2}{c}{TP4} & \multicolumn{2}{c}{TP5}
    \\\toprule
    $Q$ & $\MS$ & $\mt$ & $\MS$ & $\mt$ & $\MS$ & $\mt$
    & $\MS$ & $\mt$ & $\MS$ & $\mt$
    \\\midrule\midrule
    $A_t^{\rm OS}$ 
    & $2758$ & $2651$ & $-2525$ & --- & $940.6$ & $933.4$ & $953.0$ & $917.2$ & $-87.80$ & $-25.83$
    \\\midrule
    $M_{\tilde{Q}_{3}}^{\rm OS}$ 
    & $1507$ & $1181$ & $2542$ & --- & $1003$ & $856.2$ & $750.4$ & $655.5$ & $1547$ & $1494$
    \\\midrule
    $M_{\tilde{t}_{\rm R}}^{\rm OS}$ 
    & $1507$ & $1055$ & $513$ & --- & $1003$ & $822.2$ & $750.4$ & $629.5$ & $1547$ & $1460$
    \\\midrule\midrule
    $\MHp$ & 2759 & 2755 & 6373 & --- & 641.6 & 642.1 & 455.7 & 455.4 & 615.4 & 617.3
    \\\midrule
    $m_{\tilde{t}_1}^{\rm OS}$ & $1355$ & $904.3$ & $507.0$ & --- & $939.9$ & $761.7$ & $667.2$ & $547.0$
    & $1548$ & $1469$
    \\\midrule
    $m_{\tilde{t}_2}^{\rm OS}$ & $1661$ & $1321$ & $2554$ & --- & $1088$ & $940.9$ & $859.4$ & $764.2$
    & $1565$ & $1504$
    \\\bottomrule
  \end{tabular}
  \caption{OS parameters in the TP  scenarios obtained by the routines
    of \NC. Here $A_t^{\OS}$  is the trilinear soft-breaking parameter
    in   the   stop    sector,   while   $M_{\tilde{Q}_3}^{\OS}$   and
    $M_{\tilde{t}_R}^{\OS}$ are the soft SUSY-breaking mass parameters
    in   the  stop-sector   for   the  \SUL\   doublet  and   singlet,
    respectively.   $\MHp$ denotes  the  OS renormalized  mass of  the
    charged Higgs boson, and  $\mstop{i}^{\OS}$ denotes the obtained OS
    masses of the two top squarks.  All parameter values are given in
    GeV.}
  \label{tab:OS}
\end{table}

\begin{figure}[htb!]
 \centering
 \includegraphics{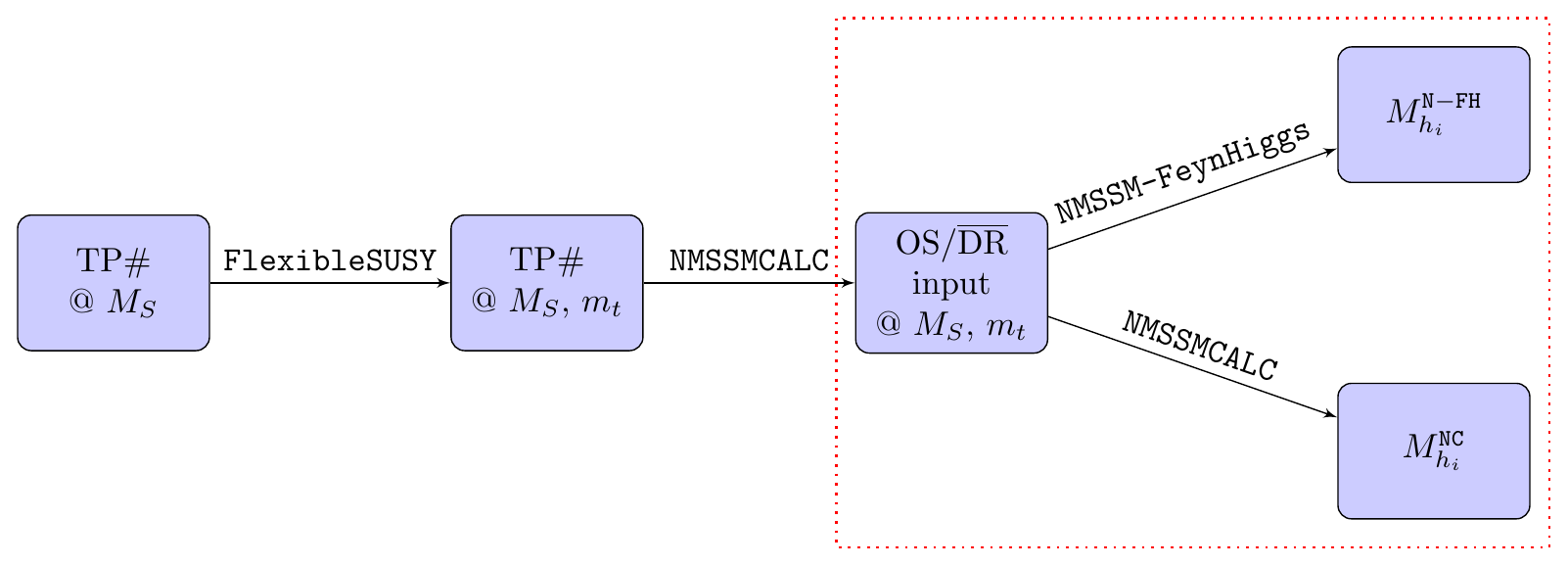}
 \caption{Steps  performed to  obtain predictions  for the  masses and
   mixing matrices of the \cp-even Higgs  fields at the scale $Q$. The
   red dashed line denotes the actual calculation of the Higgs masses.
   It is independent of the RGE evolution and the OS conversion of the
   input parameters.}
\label{fig:Alg}
\end{figure}

The current version of \NFH\ requires an OS mass for the charged Higgs
boson  as   input.  In  order   to  obtain  this  quantity   from  the
\DRbar\  parameters  specified  in \refta{tab:TPDefinition},  we  used
routines implemented in \NC. Those  routines calculate a two-loop pole
mass for the  charged Higgs boson from the given  input parameters. We
use the \DRbar\ option for  the renormalization of the top/stop sector
for this computation.   The result is treated as the  OS mass for this
comparison and  is used as  an input value  for both codes,  \NFH\ and
\NC.%
\footnote{The choice  of using  \DRbar\ stop-sector parameters  in the
  evaluation  of $\MHp$  (and  not  the ones  converted  to OS)  later
  facilitates the comparison with a pure \DRbar\ calculation.}%
The OS shifted stop-sector parameters and the resulting stop masses as
obtained with  the routines  of \NC\ are  given in  \refta{tab:OS}. An
overview of  the procedure  of how the  Higgs-boson masses  and mixing
matrices are  obtained from the  original definition of  the scenarios
TP1--TP5  is   given  in  \reffi{fig:Alg}.   The   transition  between
\DRbar\ and OS parameters can give  rise to significant shifts in case
some of the  involved SUSY masses are heavy, see  e.g.\ the discussion
in refs.~\cite{Carena:2000dp,Braathen:2016mmb}.

%% file: sections/calculation.tex
\begin{table}[h]
  \scriptsize \centering
  \begin{tabular}{llcccccccccc}
    &
    & \multicolumn{2}{c}{TP1} & \multicolumn{2}{c}{TP2} & \multicolumn{2}{c}{TP3}
    & \multicolumn{2}{c}{TP4} & \multicolumn{2}{c}{TP5}
    \\\toprule
    & $Q$ & $\MS$ & $\mt$ & $\MS$ & $\mt$ & $\MS$ & $\mt$
    & $\MS$ & $\mt$ & $\MS$ & $\mt$
    \\\midrule\midrule
    \multirow{2}[4]{*}{$h_1$} 
    & \NCs\ \OS
    & \bf 121.84 & \bf 113.47 & \bf 120.42 & --- 
    & \it 89.92 & \it 88.81 & \bf 126.44 & \bf 126.65 & \it 119.54 & \it 117.63
    \\\cmidrule{2-12}
    & \NFHs
    & \bf 115.70 & \bf 113.20 & \bf 114.12 & ---
    & \it 89.67 & \it 89.36 & \bf 126.17 & \bf 126.29 & \it 118.47 & \it 117.95
    \\
    \midrule
    \midrule
    \multirow{2}[4]{*}{$h_2$} 
    & \NCs\ \OS
    & \it 1797.45 & \it 1797.57 & \it 5951.36 & --- 
    & \bf 126.16 & \bf 125.80 & \it 143.32 & \it 142.73 & \bf 124.44 & \bf 123.51
    \\\cmidrule{2-12}
    & \NFHs
    & \it 1797.45 & \it 1797.62 & \it 5951.36 & --- 
    & \bf 124.55 & \bf 125.02 & \it 143.11 & \it 142.68 & \bf 122.93 & \bf 123.10
    \\
    \midrule
    \midrule
    \multirow{2}[4]{*}{$h_3$} 
    & \NCs\ \OS
    & 2755.73 & 2752.14 & 6370.77 & --- & 652.60 & 652.70 & 467.89 & 467.35 & 627.18 & 628.72
    \\\cmidrule{2-12}
    & \NFHs
    & 2755.79 & 2752.25 & 6370.85 & --- & 652.17 & 652.65 & 467.10 & 467.33 & 626.59 & 628.76
    \\\bottomrule
  \end{tabular}
 \caption{Mass predictions  for the  \cp-even scalars for  TP1--5 when
   using  the indicated  versions of  \NC\ and  \NFH\ as  specified in
   \refse{sec:outofthebox}.  The mass values of the SM-like scalar are
   written  in  bold  fonts,  those  of  the  singlet-like  scalar  in
   italics.}
  \label{tab:resOB}
\end{table}

\section{Predictions for masses and mixing matrices of the \cp-even scalars}
\label{sec:Results}

In this section we analyze the  differences in the predictions for the
\cp-even Higgs boson masses and mixing matrices from \NFH\ and \NC. We
start by a comparison of  the ``out-of-the-box'' results including the
corrections of up to \order{\alt\als}, where sizeable differences show
up. In  order to understand  the origin  of these differences  we then
perform a comparison at the one-loop level, where we find that part of
the differences can be attributed  to the different renormalization of
the electroweak sector in the two codes.  This difference, which is of
the order of unknown electroweak two-loop corrections, can be adjusted
by  an appropriate  reparametrization  of the  \NFH\  result. We  then
continue with an analysis at  the two-loop level, where we investigate
the  effect of  the strong  coupling  constant and  the genuine  NMSSM
corrections to the  Higgs boson self-energies. Finally  we compare the
results   obtained   with   the   OS  version   of   \NC\   with   the
\DRbar\  calculation. The  identification  of the  various sources  of
differences   between  the   different   calculations  are   important
ingredients for a reliable estimate  of the intrinsic uncertainties in
the Higgs-boson mass and mixing-matrix calculations in the NMSSM.


\subsection{``Out-of-the-box'' results}
\label{sec:outofthebox}

In the first step of our  comparison the masses and mixing matrices of
the \cp-even Higgs-sector are evaluated  for the TP scenarios with the
``out-of-the-box'' versions of \NC\ and \NFH, restricting the two-loop
corrections in  the latter code to  $\mathcal{O}{(\alpha_t \alpha_s)}$
(cf.   \refse{sec:calcho}).   For the  results  of  \NC\ we  used  the
``out-of-the-box''  version with  the on-shell  renormalization scheme
for   the  top/stop   sector   and  the   charged   Higgs  mass   (see
\refse{sec:drbaros}), labelled  ``\NCs\ OS''.  The OS  parameters used
as   numerical  input   for  both   codes  have   been  specified   in
\refta{tab:OS}.

The  obtained   numerical  results  for   the  masses  are   given  in
\refta{tab:resOB},  the results  for the  mixing-matrix elements,  see
\refeq{Uh},  are  given  in \refta{tab:originalZH13_A}.   For  all  TP
scenarios except for TP5 we identify  the field $h_i$ with the largest
value for  \ZHabs{i2} as the SM-like  field, since it has  the largest
coupling  to the  top-quark.  We  refer to  the field  $h_i$ with  the
largest value for \ZHabs{i3} as  the singlet-like field.  For TP5 both
lighter  fields have  similar or  sizeable values  for \ZHabs{i2}  and
\ZHabs{i3}, in particular  at the scale $\mt$.  In this  case we refer
to the field $h_2$  with the mass closer to 125~GeV  as SM-like and to
the  lighter field  $h_1$ as  singlet-like (for  the two-loop  results
considered here).

%
\begin{table}[h]
 \scriptsize \centering
 \begin{tabular}{lllccccccccc}
 & & & \multicolumn{3}{c}{TP1} & \multicolumn{3}{c}{TP2} & \multicolumn{3}{c}{TP3}
 \\\toprule
 $i$ & $Q$ &
 & \ZHabs{i1} & \ZHabs{i2} & \ZHabs{i3}
 & \ZHabs{i1} & \ZHabs{i2} & \ZHabs{i3}
 & \ZHabs{i1} & \ZHabs{i2} & \ZHabs{i3}
 \\\midrule\midrule
 \multirow{4}[4]{*}{$1$} & \multirow{2}[0]{*}{$ \MS$}
 & \NCs\ \OS 
 & {0.1039} & {0.9946} & {0.0076}
 & {0.1034} & {0.9946} & {0.0004}
 & {0.2199} & {0.1994} & {0.9549}
 \\ 
 & & \NFHs
 & 0.1039 & 0.9946 & 0.0071
 & 0.1034 & 0.9946 & 0.0004
 & 0.2134 & 0.2064 & 0.9549
 \\\cmidrule{2-12}
 & \multirow{2}[0]{*}{$\mt$}
 & \NCs\ \OS
 & {0.1006} & {0.9949} & {0.0073}
 & {---} & {---} & {---} 
 & {0.2236} & {0.2210} & {0.9493}
 \\ 
 & & \NFHs 
 & 0.1006 & 0.9949 & 0.0071
 & {---} & {---} & {---}
 & 0.2245 & 0.2264 & 0.9478
 \\\midrule\midrule
 \multirow{4}[4]{*}{$2$} & \multirow{2}[0]{*}{$ \MS$}
 & \NCs\ \OS
 & {0.0075} & {0.0068} & {0.9999}
 & {0.0096} & {0.0006} & {1.}
 & {0.2797} & {0.9249} & {0.2575}
 \\ 
 & & \NFHs
 & 0.0071 & 0.0064 & 1.
 & 0.0090 & 0.0005 & 1.
 & 0.2820 & 0.9228 & 0.2625
 \\\cmidrule{2-12}
 & \multirow{2}[0]{*}{$\mt$}
 & \NCs\ \OS
 & {0.0075} & {0.0066} & {0.9999}
 & {---} & {---} & {---}
 & {0.2659} & {0.9232} & {0.2775}
 \\
 & & \NFHs
 & 0.0072 & 0.0064 & 1.
 & {---} & {---} & {---}
 & 0.2656 & 0.9216 & 0.2831
 \\\midrule\midrule
 \multirow{4}[4]{*}{$3$} & \multirow{2}[0]{*}{$\MS$}
 & \NCs\ \OS
 & {0.9946} & {0.1040} & {0.0068}
 & {0.9946} & {0.1034} & {0.0096}\
 & {0.9346} & {0.3237} & {0.1476}
 \\ 
 & & \NFHs 
 & 0.9946 & 0.1039 & 0.0064
 & 0.9946 & 0.1034 & 0.009
 & 0.9354 & 0.3253 & 0.1388
 \\\cmidrule{2-12}
 & \multirow{2}[0]{*}{$\mt$}
 & \NCs\ \OS
 & {0.9949} & {0.1006} & {0.0068}
 & {---} & {---} & {---}
 & {0.9377} & {0.3144} & {0.1476}
 \\
 & & \NFHs 
 & 0.9949 & 0.1006 & 0.0065
 & {---} & {---} & {---}
 & 0.9376 & 0.3153 & 0.1467
 \\\bottomrule\bottomrule\\
 \end{tabular}
 \\
 \begin{tabular}{lllcccccc} 
 & & & \multicolumn{3}{c}{TP4} & \multicolumn{3}{c}{TP5}
 \\\toprule
 $i$ & $Q$ &
 & \ZHabs{i1} & \ZHabs{i2} & \ZHabs{i3}
 & \ZHabs{i1} & \ZHabs{i2} & \ZHabs{i3}
 \\\midrule\midrule
 \multirow{4}[4]{*}{$1$} & \multirow{2}[0]{*}{$\MS$}
 & \NCs\ \OS
 & {0.4813} & {0.7432} & {0.4648}
 & {0.2845} & {0.3943} & {0.8738}
 \\ 
 & & \NFHs 
 & 0.4873 & 0.8022 & 0.3448
 & 0.3540 & 0.7764 & 0.5214
 \\\cmidrule{2-9}
 & \multirow{2}[0]{*}{$\mt$}
 & \NCs\ \OS
 & {0.4766} & {0.7886} & {0.3885}
 & {0.3393} & {0.6991} & {0.6294}
 \\ 
 & & \NFHs
 & 0.4788 & 0.8104 & 0.3377
 & 0.3390 & 0.7204 & 0.6051
 \\\midrule\midrule
 \multirow{4}[4]{*}{$2$} & \multirow{2}[0]{*}{$\MS$}
 & \NCs\ \OS
 & {0.0895} & {0.4858} & {0.8694}
 & {0.2224} & {0.8594} & {0.4603}
 \\ 
 & & \NFHs
 & 0.0334 & 0.3775 & 0.9254
 & 0.0564 & 0.5387 & 0.8406
 \\\cmidrule{2-9}
 & \multirow{2}[0]{*}{$\mt$}
 & \NCs\ \OS
 & {0.0411} & {0.4215} & {0.9059}
 & {0.0882} & {0.6425} & {0.7612}
 \\
 & & \NFHs
 & 0.0171 & 0.3761 & 0.9264
 & 0.0819 & 0.6181 & 0.7818
 \\\midrule\midrule
 \multirow{4}[4]{*}{$3$} & \multirow{2}[0]{*}{$\MS$}
 & \NCs\ \OS 
 & {0.8720} & {0.4600} & {0.1673}
 & {0.9325} & {0.3253} & {0.1568}
 \\ 
 & & \NFHs 
 & 0.8726 & 0.4625 & 0.1571
 & 0.9335 & 0.3270 & 0.1469
 \\\cmidrule{2-9}
 & \multirow{2}[0]{*}{$\mt$}
 & \NCs\ \OS
 & {0.8782} & {0.4477} & {0.1685}
 & {0.9365} & {0.3137} & {0.1564}
 \\ 
 & & \NFHs
 & 0.8778 & 0.4493 & 0.1662
 & 0.9372 & 0.3146 & 0.1505
 \\\bottomrule
 \end{tabular}
 \caption{Absolute  values  for  the  mixing matrix  elements  of  the
   \cp-even scalar sector for TP1--5  when using the indicated versions
   of \NC\ and \NFH\ as specified in the text, see \refse{sec:outofthebox}.}
 \label{tab:originalZH13_A}
\end{table}

\begin{figure}[h]
 \centering
 \includegraphics[width=.4\textwidth,width=.33\textheight]{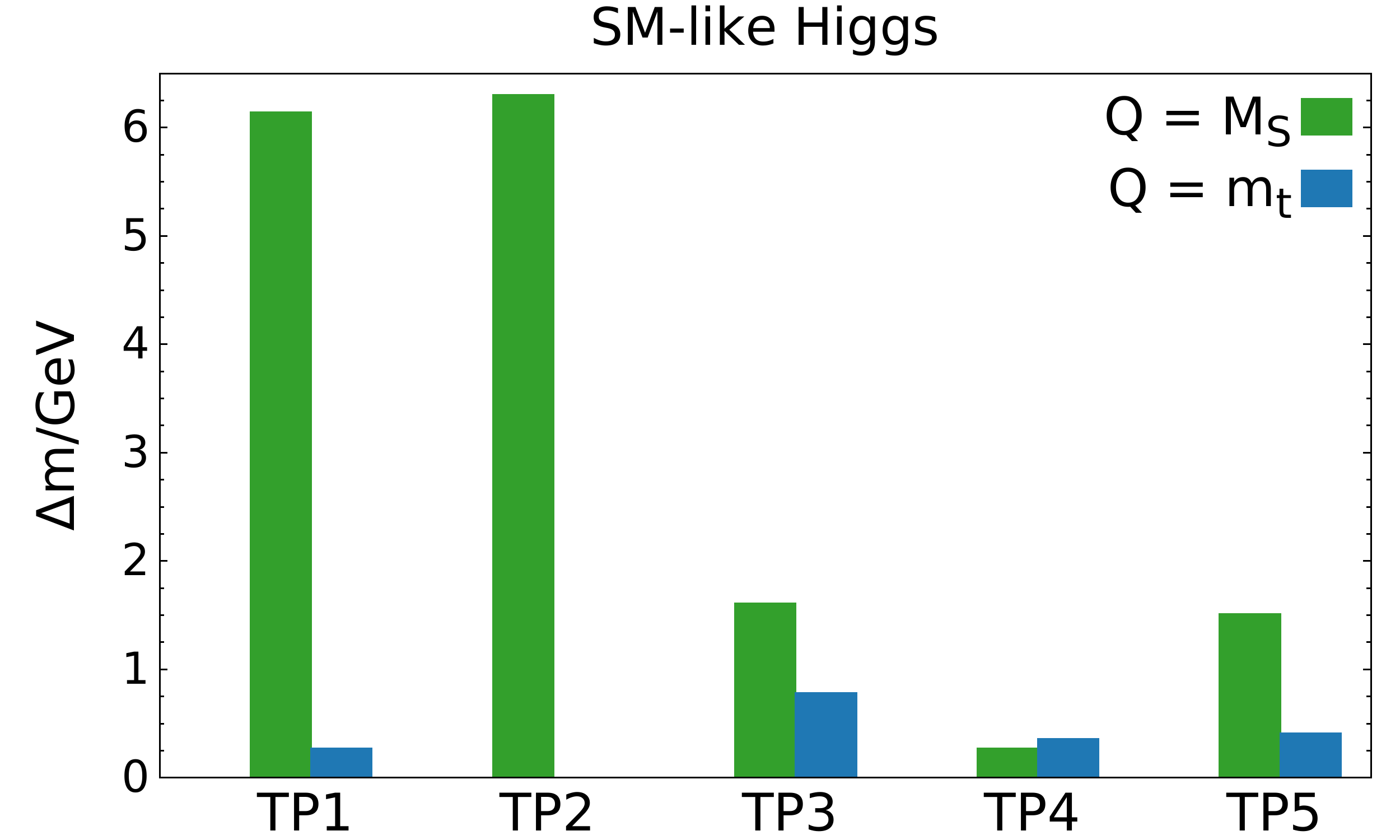}
 \;\;
 \includegraphics[width=.4\textwidth,width=.33\textheight]{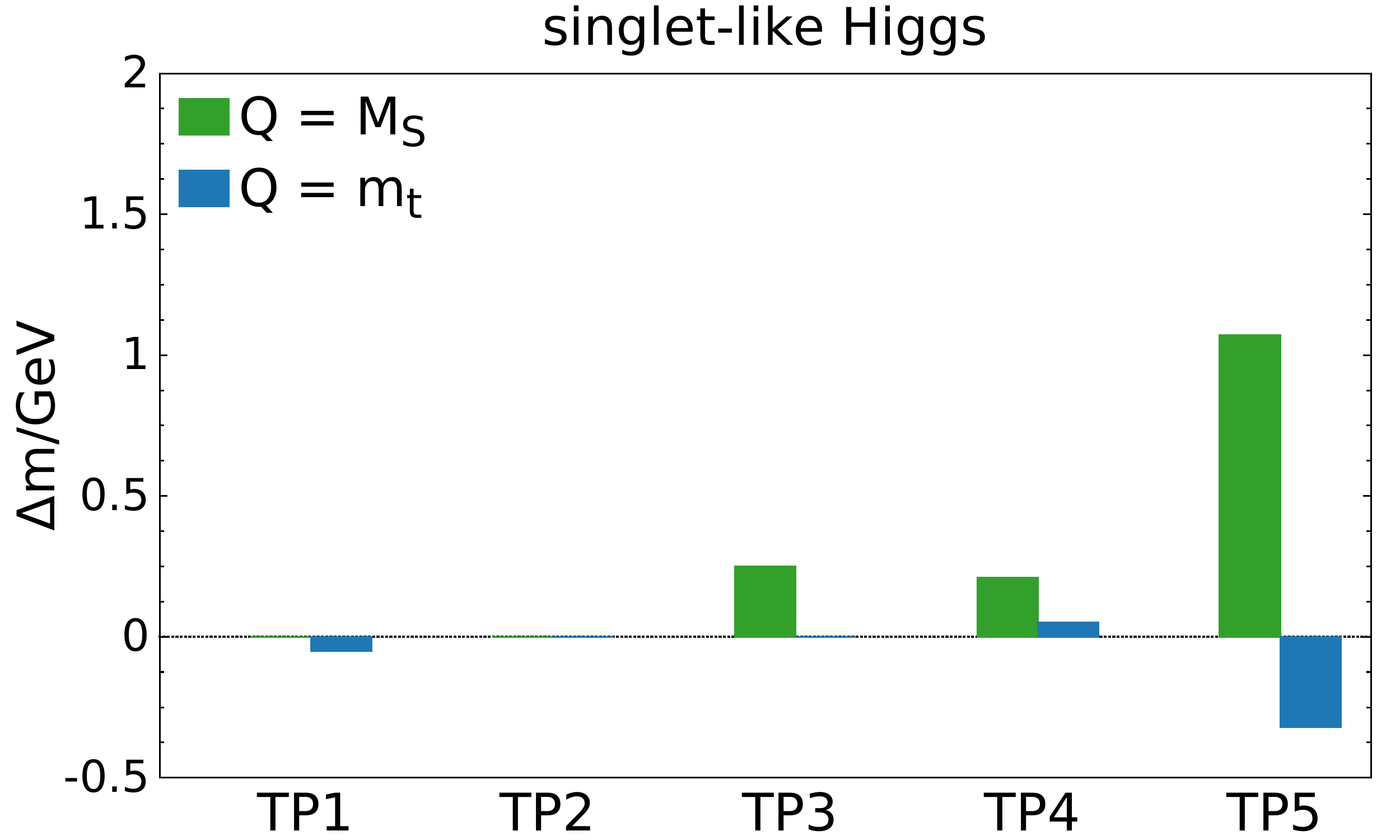}
 \\
 \includegraphics[width=.4\textwidth,width=.33\textheight]{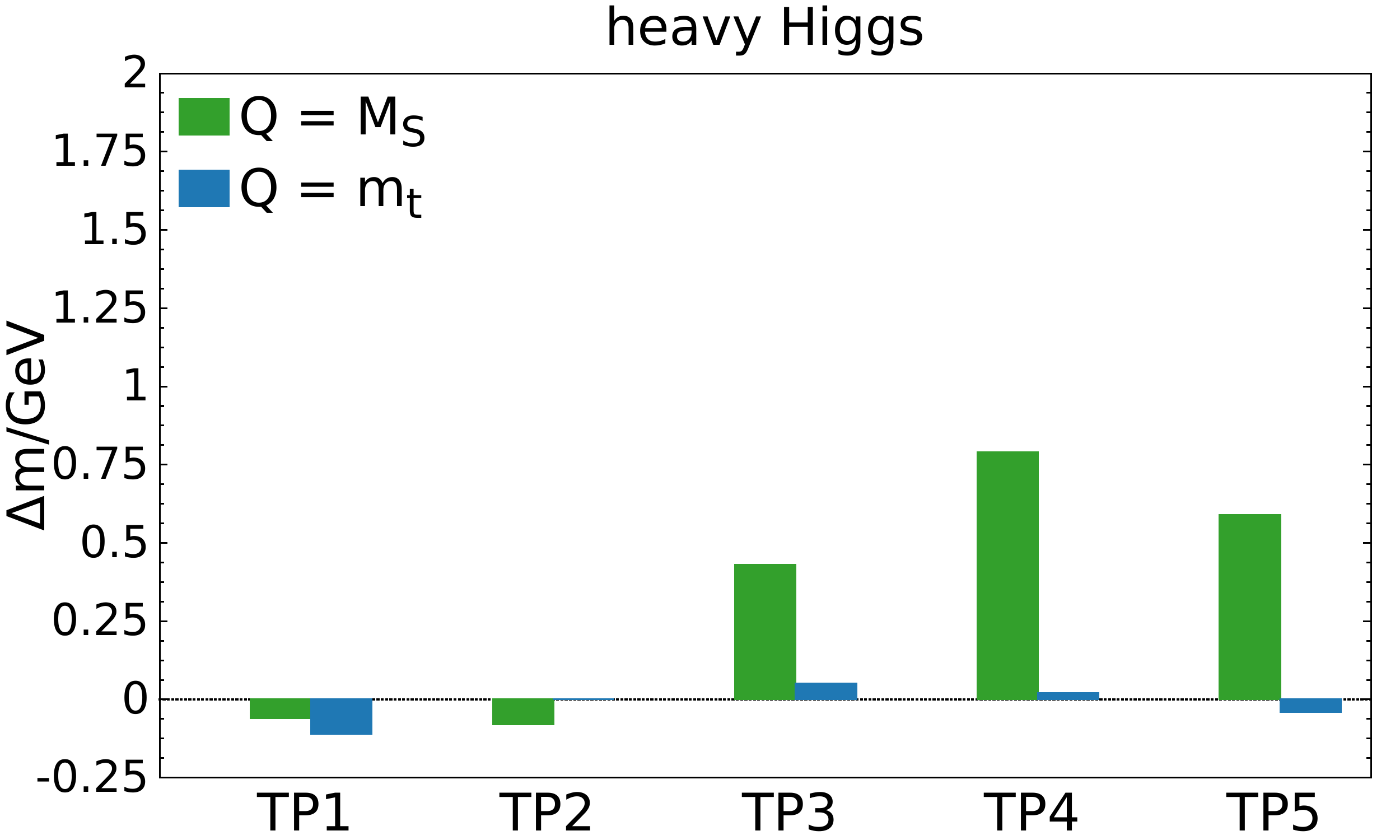}
 \caption{Difference  $\Delta{m} =  M_h^{\NCs} -  M_h^{\NFHs}$ between
   the predicted Higgs  masses of the SM-like,  singlet-like and heavy
   Higgs  at the  two-loop level  calculated  at the  scales \MS\  and
   \MTO.}
 \label{fig:tot}
\end{figure}

In   \reffi{fig:tot}  the   difference  $\Delta{m}   =  M_h^{\NCs}   -
M_h^{\NFHs}$  between the  mass  predictions obtained  with the  codes
``\NCs\ OS''  and ``\NFHs'' are  shown. All  values are given  for two
input scales  $Q$, the SUSY  mass scale $Q  = \MS$ in  lighter (green)
bars and  the top-quark mass  scale $Q=  \MTO = 172.9~\gev$  in darker
(blue) bars.

When comparing the  Higgs masses obtained with \NFH\ and  \NC, we find
that for the SM-like field the difference between the mass predictions
of the two codes can be larger  than 6~\gev\ for the scenarios TP1 and
TP2 and  the evaluation  at the high  scale $M_S$.   These differences
between the mass predictions obtained  at the two-loop level appear to
be unusually large.  The origin of these differences will be addressed
in  the following  sections.  For  the  evaluations at  the low  scale
$\mt$,  however,  the differences  between  the  codes do  not  exceed
1~\gev\ for all scenarios (as mentioned above, the scenario TP2 is not
evaluated at the  scale \MTO\ in our numerical analysis,  as it yields
tachyonic  stop-masses).  The  difference  between the  masses of  the
heaviest fields,  which are  always doublet-like in  all TP-scenarios,
remain rather small with less than  0.8~\gev, which is at the permille
level for the considered scenarios.  For scenarios with a singlet-like
field that is heavier than the  SM-like field, i.e.\ TP1, TP2 and TP4,
the  absolute   difference  between  the  mass   predictions  for  the
singlet-like  field  remains  below  0.25~\gev, while  for  a  lighter
singlet-field the differences can be as large as 1.1~\gev.  The mixing
matrix elements of  the SM-like Higgs boson obtained by  the two codes
agree  within  $\approx  2\%$  for  the  scenarios  TP1--3.   For  the
singlet-like  Higgs  boson  the   differences  in  the  mixing  matrix
elements, on the other hand, can be substantial in these scenarios. In
case of  the heavy Higgs $h_3$  the matrix elements differ  by at most
10\%. For  TP4 and TP5  we find good  agreement for most  entries, but
larger  differences, up  to a  factor of  four, can  occur for  matrix
elements  that are  themselves small.   In general,  we find  that the
differences for  the mixing matrix elements  seem to be by  far larger
than for  the mass predictions,  where the relative  differences never
exceed 6\%.  This is in particular the  case for TP4 and TP5, where we
have large singlet  admixtures to the SM-like Higgs field  with a mass
around  125~\gev.  In these  scenarios  the  Higgs masses  and  mixing
matrices appear  to be very  sensitive to relatively small  changes of
the parameters.  We remark that this can lead to different conclusions
whether a parameter point is excluded or not by LHC data.

In  the following  we will  analyze the  observed differences  in more
detail. We will  start with a discussion of the  chosen test-point TP1
and a study of the differences at the one-loop level.


\subsection{The test-point TP1}

In  \refta{tab:resOB}   very  large   differences  between   the  mass
prediction in the  MSSM-like scenario TP1 can be  found when comparing
the result for the SM-like \cp-even Higgs field obtained with the same
code  at either  the scale  $\MS$ or  \MTO. They  can be  as large  as
$\approx  8.5~\gev$ for  the  SM-like scalar  with  \NC\ and  $\approx
3~\gev$ for the  heavy scalar with both codes (it  should be noted, of
course,  that for  the heavy  scalar this  amounts to  a much  smaller
relative effect than for the  SM-like scalar). These different results
at  two  different  scales  may seem  surprising  since  the  physical
situation  before and  after the  evolution of  the parameters  of the
scenario should  be identical. Furthermore,  we observed for  TP1 that
changing the  renormalization scheme from  an OS renormalization  to a
\DRbar\ renormalization in the top/stop  sector for TP1 at $Q=m_t$, as
in \refse{sec:DRbar}, changes the mass  of the lightest Higgs boson by
8.7 GeV.%
\footnote{  In  \cite{Braathen:2016mmb}  a  similar  effect  has  been
  observed.  It  was   found  that  in  case  of   large  gluino  mass
  $m_{\tilde{g}}$ OS renormalization of  the top/stop sector (with the
  physical  stop  masses   used  as  input)  is   advantageous  over  a
  \DRbar\   renormalization  of   the   top/stop   sector  since   the
  \DRbar\    renormalization    leads    to    terms    enhanced    in
  $m_{\tilde{g}}^2/\MS^2$.}  It  can be seen from  \refta{tab:OS} that
the running  from $\MS$  to \MTO,  which we  have performed  with {\tt
  FlexibleSUSY}, and the conversion from the \DRbar\ to the OS scheme,
which we have  carried out as described  in \refse{sec:drbaros}, gives
rise to large shifts of the OS  parameters of the scalar top sector in
this scenario. These large effects  are induced by the large splitting
between the  gluino-mass parameter $M_3  = 3~{\rm TeV}$ and  the other
parts      of       the      spectrum      in       this      scenario
(cf.\  tab.~\ref{tab:TPDefinition}).  In  such  a  case  a  consistent
decoupling of the heavy gluino should be performed (see the discussion
in ref.~\cite{Muhlleitner:2008yw}),  which is beyond the  scope of our
present analysis. If the heavy gluino  is kept in the spectrum for the
running from $\MS$ to \MTO\ and for the conversion from the \DRbar\ to
the  OS  scheme, the  obtained  low-scale  scenario corresponds  to  a
different physical  situation than the high-scale  one. The comparison
of  the results  of the  high-scale and  the low-scale  scenario would
therefore not describe the difference between the two calculations but
rather  a difference  between  two distinct  physical situations.   We
found also  that the  parameters obtained at  \MTO\ define  a scenario
that is highly sensitive to variations of the stop parameters. This is
in particular true for $A_t$, which  at the input scale $M_S$ is large
compared to the  other soft-breaking parameters of  the scenario.  For
example, at the low scale  \MTO , changes of $\left|\Delta{A_t}\right|
\approx 100~\gev$ can yield a change of the mass of the lightest Higgs
of $|\Delta{m_{h_1}}| \approx 2~\gev$.  Since the observed effects are
independent    of    the    Higgs-mass   calculation    itself    (see
\reffi{fig:Alg}), we  regard the scenario  TP1 in the present  form as
not suitable for the discussion  of theoretical uncertainties. We will
therefore omit this scenario in  the following. Note, however, that we
explicitly  checked that  by doing  the  adjustments of  the codes  as
described  in  the following  sections  we  find very  good  agreement
between  \NFH\ and  \NC\ for  both  the low-scale  and the  high-scale
scenario  of TP1.  The results  for the  low-scale and  the high-scale
scenario differ  by about 3~GeV  for the  SM-like Higgs in  both codes
after the adjustments.


\subsection{Renormalization of the electromagnetic coupling constant \boldmath{$\alpha$}}
\label{sec:1Lreparametrization}

In order  to disentangle the  effects arising from differences  in the
renormalization,   the  coupling   constants   and  the   higher-order
corrections,  we  start by  comparing  the  one-loop results  for  the
Higgs-boson  masses and  mixing  matrices. Here  we  use the  versions
'\NCs\  OS' and  '\NFHs' as  described  in the  previous section,  but
restrict  the predictions  to the  pure one-loop  contribution to  the
Higgs-boson self-energies.   In this case the  only difference between
the calculations stems from the different renormalization prescription
of    the   electromagnetic    coupling    constant   $\alpha$,    see
\refta{tab:DiffNcNfhVanilla}.  The corresponding numerical results for
the \cp-even Higgs-boson masses are given in \refta{tab:resOB1L}.

\begin{table}[htb!]
 \scriptsize \centering
 \begin{tabular}{llcccccccc}
 &
 & \multicolumn{2}{c}{TP2} & \multicolumn{2}{c}{TP3}
 & \multicolumn{2}{c}{TP4} & \multicolumn{2}{c}{TP5}
 \\\toprule
 & $Q$ 
 & $\MS$ & $\mt$ & $\MS$ & $\mt$
 & $\MS$ & $\mt$ & $\MS$ & $\mt$
 \\\midrule\midrule
 \multirow{3}[6]{*}{$h_1$} 
 & \NCs\ \OS
 & \bf 140.67 & ---
 & \it 90.47 & \it 90.12 & \bf 132.48 & \bf 134.10 & \it 120.93 & \it 120.57
 \\\cmidrule{2-10}
 & \NFHs
 & \bf 139.68 & ---
 & \it 90.30 & \it 89.98 & \bf 132.96 & \bf 133.42 & \it 120.82 & \it 120.51
 \\\cmidrule{2-10}
 & \NFHs\ $\alpha(\MZ)$
 & \bf 140.67 & ---
 & \it 90.38 & \it 89.96 & \bf 132.73 & \bf 133.22 & \it 120.91 & \it 120.53
 \\
 \midrule
 \midrule
 \multirow{3}[6]{*}{$h_2$} 
 & \NCs\ \OS
 & \it 5951.36 & --- 
 & \bf 136.35 & \bf 138.25 & \it 146.45 & \it 146.84 & \bf 135.56 & \bf 138.27
 \\\cmidrule{2-10}
 & \NFHs
 & \it 5951.36 & --- 
 & \bf 136.73 & \bf 137.04 & \it 146.82 & \it 146.05 & \bf 136.30 & \bf 137.06
 \\\cmidrule{2-10}
 & \NFHs\ $\alpha(\MZ)$
 & \it 5951.36 & --- 
 & \bf 136.74 & \bf 137.19 & \it 146.59 & \it 145.89 & \bf 136.32 & \bf 137.26
 \\
 \midrule
 \midrule
 \multirow{3}[6]{*}{$h_3$} 
 & \NCs\ \OS
 & 6370.75 & --- & 652.80 & 653.13 & 468.63 & 468.32 & 627.34 & 629.19
 \\\cmidrule{2-10}
 & \NFHs
 & 6370.83 & --- & 652.49 & 653.04 & 468.01 & 468.30 & 626.94 & 629.14
 \\\cmidrule{2-10}
 & \NFHs\ $\alpha(\MZ)$
 & 6370.88 & --- & 652.62 & 653.18 & 468.37 & 468.60 & 627.08 & 629.23
 \\\bottomrule
 \end{tabular}
 \caption{One-loop  mass  predictions  for the  \cp-even  scalars  for
   TP2--5  when using  the indicated  versions  of \NC\  and \NFH\  as
   specified in  sec.~\ref{sec:1Lreparametrization}.  The  mass values
   of  the SM-like  scalar are  written in  bold fonts,  those of  the
   singlet-like scalar in italics.}
 \label{tab:resOB1L}
\end{table}

\begin{table}[htb!]
 \scriptsize \centering
 \begin{tabular}{lllcccccc} 
 & & & \multicolumn{3}{c}{TP2} & \multicolumn{3}{c}{TP3}
 \\\toprule
 $i$ & $Q$ &
 & \ZHabs{i1} & \ZHabs{i2} & \ZHabs{i3}
 & \ZHabs{i1} & \ZHabs{i2} & \ZHabs{i3}
 \\\midrule\midrule
 \multirow{4}[4]{*}{$1$} & \multirow{2}[0]{*}{$\MS$}
 & \NCs\ \OS
 & {0.1034} & {0.9946} & {0.0004}
 & {0.2048} & {0.1503} & {0.9672} 
 \\ 
 & & \NFHs\ $\alpha{(\MZ)}$ 
 & 0.1034 & 0.9946 & 0.0004
 & 0.1981 & 0.1443 & 0.9695
 \\\cmidrule{2-9}
 & \multirow{2}[0]{*}{$\mt$}
 & \NCs\ \OS
 & {---} & {---} & {---}
 & {0.2048} & {0.1553} & {0.9664} 
 \\ 
 & & \NFHs\ $\alpha{(\MZ)}$
 & {---} & {---} & {---}
 & 0.2054 & 0.1601 & 0.9655
 \\\midrule\midrule
 \multirow{4}[4]{*}{$2$} & \multirow{2}[0]{*}{$\MS$}
 & \NCs\ \OS
 & {0.0096} & {0.0006} & {1.0000}
 & {0.2935} & {0.9332} & {0.2071}
 \\ 
 & & \NFHs\ $\alpha{(\MZ)}$
 & 0.0092 & 0.0005 & 1.0000
 & 0.2971 & 0.9337 & 0.1997
 \\\cmidrule{2-9}
 & \multirow{2}[0]{*}{$\mt$}
 & \NCs\ \OS
 & {---} & {---} & {---}
 & {0.2836} & {0.9356} & {0.2104} 
 \\
 & & \NFHs\ $\alpha{(\MZ)}$
 & {---} & {---} & {---}
 & 0.2837 & 0.9344 & 0.2153
 \\\midrule\midrule
 \multirow{4}[4]{*}{$3$} & \multirow{2}[0]{*}{$\MS$}
 & \NCs\ \OS 
 & {0.9946} & {0.1034} & {0.0096}
 & {0.9338} & {0.3263} & {0.1470} 
 \\ 
 & & \NFHs\ $\alpha{(\MZ)}$ 
 & 0.9946 & 0.1034 & 0.0092
 & 0.9341 & 0.3276 & 0.1421
 \\\cmidrule{2-9}
 & \multirow{2}[0]{*}{$\mt$}
 & \NCs\ \OS
 & {---} & {---} & {---}
 & {0.9368} & {0.3172} & {0.1475} 
 \\ 
 & & \NFHs\ $\alpha{(\MZ)}$
 & {---} & {---} & {---}
 & 0.9366 & 0.3182 & 0.1466
 \\\bottomrule\bottomrule\\
 \end{tabular}
 \begin{tabular}{lllcccccc} 
 & & & \multicolumn{3}{c}{TP4} & \multicolumn{3}{c}{TP5}
 \\\toprule
 $i$ & $Q$ &
 & \ZHabs{i1} & \ZHabs{i2} & \ZHabs{i3}
 & \ZHabs{i1} & \ZHabs{i2} & \ZHabs{i3}
 \\\midrule\midrule
 \multirow{4}[4]{*}{$1$} & \multirow{2}[0]{*}{$\MS$}
 & \NCs\ \OS
 & {0.4393} & {0.5717} & {0.6929}
 & {0.2079} & {0.1294} & {0.9695} 
 \\ 
 & & \NFHs\ $\alpha{(\MZ)}$ 
 & 0.4446 & 0.5948 & 0.6697
 & 0.1998 & 0.1211 & 0.9723
 \\\cmidrule{2-9}
 & \multirow{2}[0]{*}{$\mt$}
 & \NCs\ \OS
 & {0.4422} & {0.6086} & {0.6588}
 & {0.2117} & {0.1484} & {0.9660} 
 \\ 
 & & \NFHs\ $\alpha{(\MZ)}$
 & 0.4538 & 0.6467 & 0.6131
 & 0.2151 & 0.1624 & 0.9630
 \\\midrule\midrule
 \multirow{4}[4]{*}{$2$} & \multirow{2}[0]{*}{$\MS$}
 & \NCs\ \OS
 & {0.2285} & {0.6749} & {0.7017}
 & {0.2983} & {0.9356} & {0.1889}
 \\ 
 & & \NFHs\ $\alpha{(\MZ)}$
 & 0.2188 & 0.6529 & 0.7252
 & 0.3026 & 0.9362 & 0.1788
 \\\cmidrule{2-9}
 & \multirow{2}[0]{*}{$\mt$}
 & \NCs\ \OS
 & {0.1982} & {0.6500} & {0.7336}
 & {0.2832} & {0.9367} & {0.2060} 
 \\
 & & \NFHs\ $\alpha{(\MZ)}$
 & 0.1730 & 0.6110 & 0.7725
 & 0.2812 & 0.9340 & 0.2203
 \\\midrule\midrule
 \multirow{4}[4]{*}{$3$} & \multirow{2}[0]{*}{$\MS$}
 & \NCs\ \OS 
 & {0.8688} & {0.4666} & {0.1658}
 & {0.9315} & {0.3285} & {0.1559} 
 \\ 
 & & \NFHs\ $\alpha{(\MZ)}$
 & 0.8686 & 0.4689 & 0.1602
 & 0.9319 & 0.3300 & 0.1504
 \\\cmidrule{2-9}
 & \multirow{2}[0]{*}{$\mt$}
 & \NCs\ \OS
 & {0.8747} & {0.4550} & {0.1668}
 & {0.9354} & {0.3172} & {0.1563} 
 \\ 
 & & \NFHs\ $\alpha{(\MZ)}$
 & 0.8742 & 0.4566 & 0.1654
 & 0.9352 & 0.3182 & 0.1552
 \\\bottomrule
 \end{tabular}
 \label{tab:Mixing1L}
 \caption{One-loop mixing matrix elements for TP2--5 when using the
 indicated versions of \NC\ and \NFH\ as specified in the text, see \refse{sec:1Lreparametrization}.}
\end{table}

\begin{figure}[htb!]
 \centering
 \includegraphics[width=.4\textwidth,width=.33\textheight]{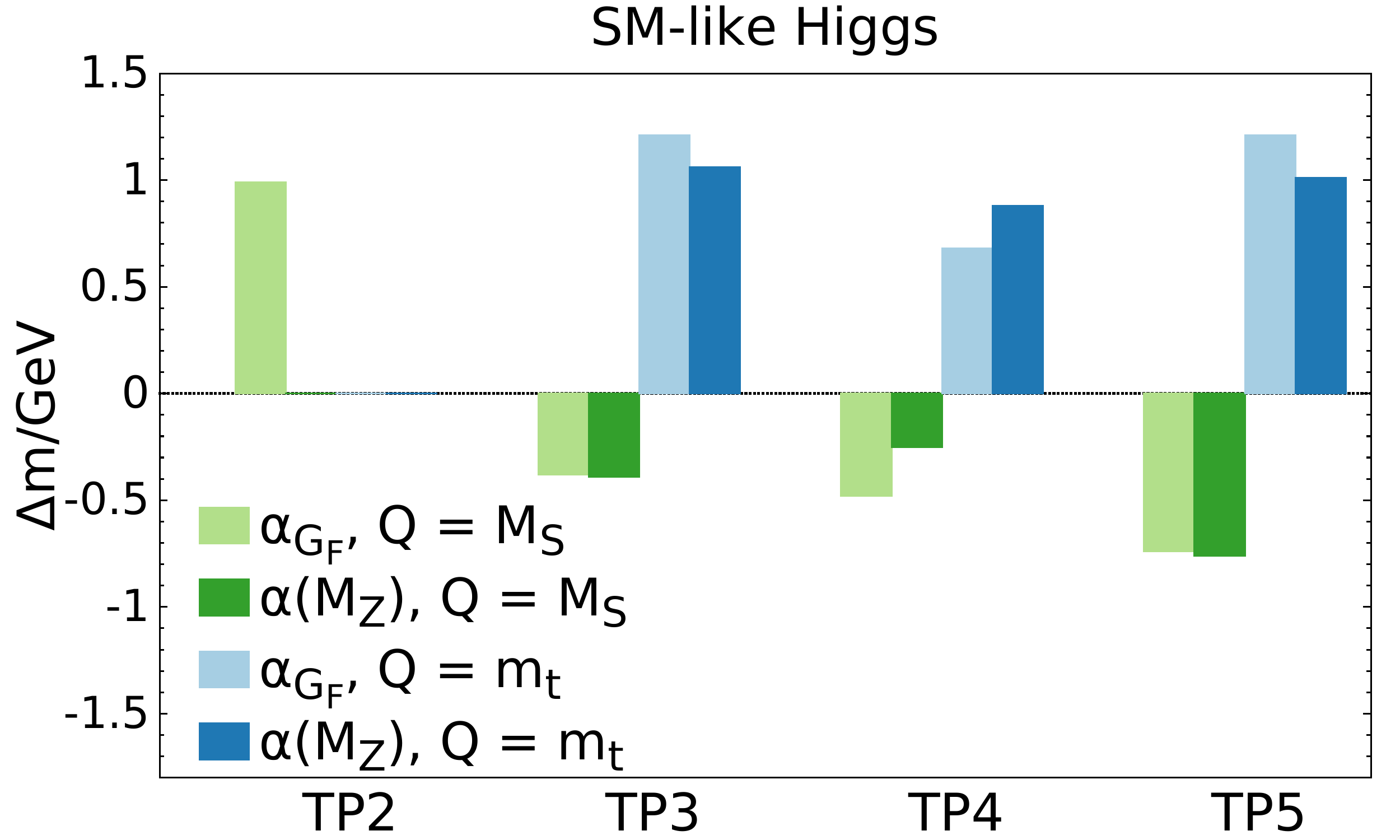}
 \;\;
 \includegraphics[width=.4\textwidth,width=.33\textheight]{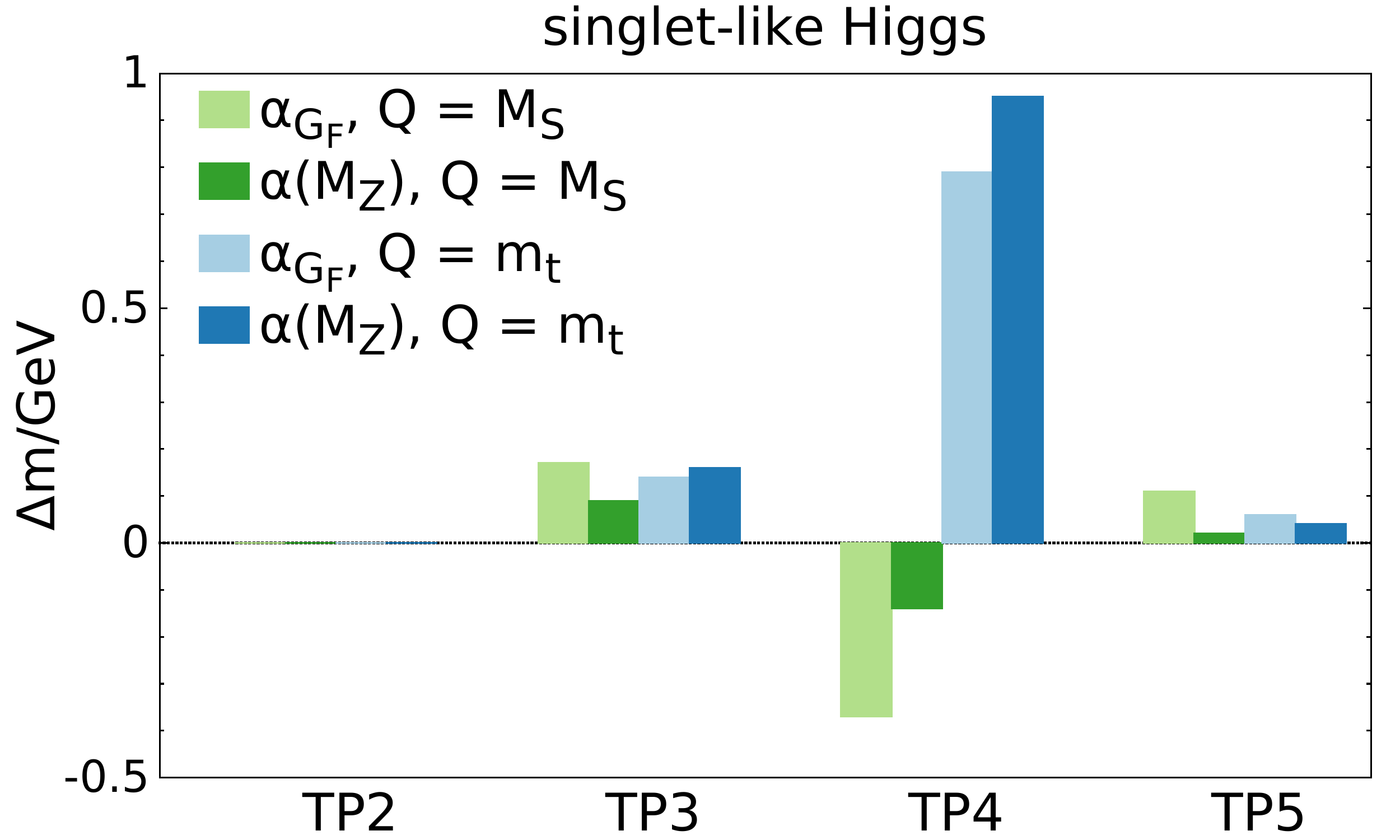}
 \\
 \includegraphics[width=.4\textwidth,width=.33\textheight]{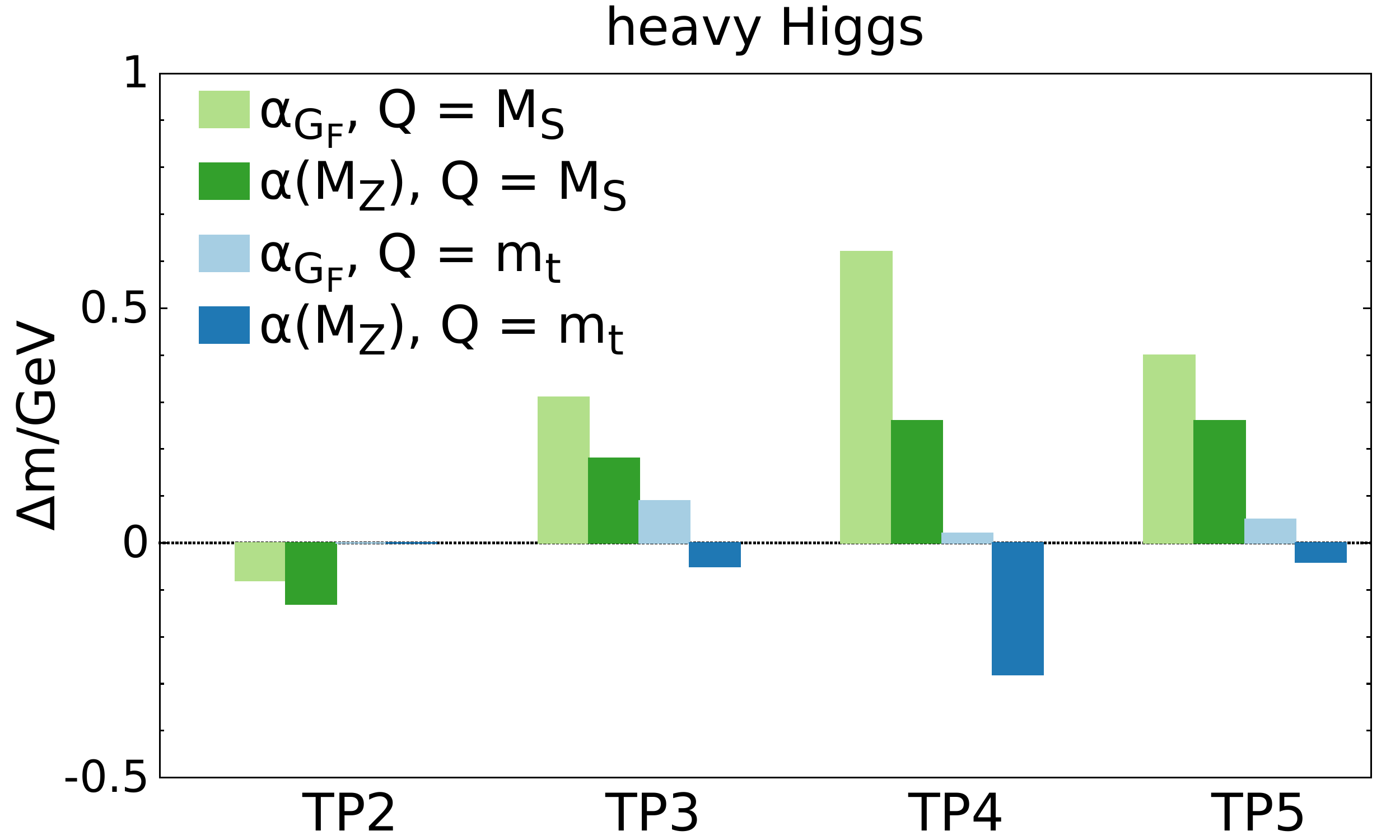}
 \caption{One-loop difference $\Delta m = M_h^{\NCs} - M_h^{\NFHs}$ at
   the  scales  $\MS$  (green)  and  $\mt$  (blue)  between  \NC\  and
   \NFH\  with the  reparametrization to  $\alpha_{G_F}$ (bright)  and
   $\alpha{(\MZ)}$ (dark).}
 \label{fig:resOB1L}
\end{figure}

The differences  between the  one-loop results applying  the different
definitions  for  the  electromagnetic  coupling  constant,  shown  in
\reffi{fig:resOB1L} by light  green bars, are smaller  compared to the
observed  differences  at  the  two-loop  level.   They  never  exceed
1.0~\gev\ for  all masses in all  scenarios.  In order to  account for
this  well-understood  difference   between  the  two  renormalization
schemes for $\alpha_{G_F}$ and $\alpha{(\MZ)}$,  which is of the order
of unknown electroweak two-loop corrections and can therefore serve as
an  indication   of  the   possible  size  of   remaining  theoretical
uncertainties of this type, we now  employ a modification of \NFH.  In
\NFH\ the treatment of $\alpha$ is  a two-step procedure: in the first
step a \DRbar\ reparametrization  for the vacuum expectation-value $v$
is applied.  In the second step  this result is then reparametrized in
terms of  a suitably chosen  expression for $\alpha$. For  the results
discussed so  far, the electric  charge is  expressed in terms  of the
Fermi constant $G_F$, the default value in \NFH.  As discussed before,
this is done to ensure that in \NFH\ the MSSM limit exactly reproduces
the   MSSM   result  of   \FH.    For   the  discussed   results   the
reparametrization of the electromagnetic coupling is only necessary up
to   the   one-loop  level,   since   the   two-loop  corrections   of
$\mathcal{O}{(\alpha_t   \alpha_s)}$  have   been   obtained  in   the
gauge-less limit (cf.~ref.~\cite{Drechsel:2016jdg}).  In the following
we adjust the  second step in the outlined procedure,  where we choose
to express the electric charge  in \NFH\ by its value $\alpha{(M_Z)}$,
the  default value  in  \NC.   This modified  version  is labelled  as
``\NFHs\ $\alpha{(\MZ)}$''.  This modification  is expected to yield a
better,  yet  not  perfect  agreement  between  the  two  codes.   The
remaining     difference      between     results      obtained     by
``\NFHs\   $\alpha{(\MZ)}$'',    where   the   electric    charge   is
reparametrized to  the value $\alpha{(M_Z)}$, and  ``\NCs\ OS'', where
the  electric charge  is  renormalized to  the value  $\alpha{(M_Z)}$,
consists  formally also  of  electroweak corrections  of two-loop  and
higher orders.

The mass  predictions for  the version ``\NFHs\  $\alpha{(\MZ)}$'' are
given  in  \refta{tab:resOB1L}.   When  compared  to  the  results  of
``\NCs\ OS''  at the scale $\MS$  the results of the  adjusted version
``\NFHs\ $\alpha{(\MZ)}$'' are in better  or equally well agreement as
the results of the previous  version ``\NFHs'' without the adjustment.
This  can be  seen  by comparing  the  light and  dark  green bars  in
\reffi{fig:resOB1L}.  For the comparison at  $Q = \MTO$, shown as blue
bars  in \reffi{fig:resOB1L},  for  the SM-like  Higgs  boson also  an
improvement  is achieved  by the  reparametrization of  $\al$ for  all
scenarios except for TP4, where  the agreement is slightly worse.  For
the  other two  Higgs bosons  the result  is less  conclusive. In  all
scenarios but  TP4 the reparametrization to  $\alpha{(\MZ)}$ yields an
improved  or equally  well agreement  as  the results  of the  version
``\NFHs''  without  the  adjustment  at both  scales.   The  mentioned
two-loop  and higher-order  effects  from  the charge  renormalization
appear to  be more important in  the scenario TP4.  The  mixing matrix
elements, see  \refta{tab:Mixing1L}, obtained  by the two  codes agree
within $\approx~10\%$  with the  largest differences occuring  for the
scenarios TP4 and TP5.  For the scenario TP4 and the two lighter Higgs
states  we  obtained  similar,  sizeable  values  for  \ZHabs{i2}  and
\ZHabs{i3} with either code at both scales $M_S$ and $\mt$, making the
assignment  of the  singlet-  and SM-like  field  ambiguous.  We  thus
follow  the  identification  obtained with  the  two-loop  calculation
described  in \refse{sec:outofthebox}.   In order  to verify  that the
observed   differences  between   the   versions   ``\NCs\  OS''   and
``\NFHs\  $\alpha{(\MZ)}$''  are  indeed  explained  by  two-loop  and
higher-order effects from the reparametrization procedure, we compared
the  predictions of  the  two versions  in the  MSSM-limit  of the  TP
scenarios.  In the MSSM-limit, with  $\lambda = \kappa \rightarrow 0$,
the  renormalization constant  of $\alpha$  drops out  as well  as the
reparametrization.  We found that in  the MSSM-limit there is complete
agreement between  the two  codes at the  expected level  of numerical
accuracy.

The  reparametrization to  $\alpha{(\MZ)}$ in  \NFH\ overall  yields a
better agreement with \NC.  The effect  on the mass prediction for the
SM-like  Higgs,  however, is  much  smaller  than  some of  the  large
differences   observed   for   the   two-loop   mass   prediction   in
\reffi{fig:tot}.   In the  subsequent  sections  a comparison  between
\NC\  and   \NFH,  where  in   the  latter  code  $\alpha$   has  been
reparametrized to  $\alpha{(\MZ)}$, will be performed  at the two-loop
level in order to identify the  differences that are not caused by the
renormalization of the electromagnetic coupling constant $\alpha$.

\begin{table}[htb!]
 \scriptsize \centering
 \begin{tabular}{llcccccccccc}
 &
 & \multicolumn{2}{c}{TP2} & \multicolumn{2}{c}{TP3}
 & \multicolumn{2}{c}{TP4} & \multicolumn{2}{c}{TP5}
 \\\toprule
 &$Q$ 
 & $\MS$ & $\mt$ & $\MS$ & $\mt$
 & $\MS$ & $\mt$& $\MS$ & $\mt$
 \\\midrule\midrule
 \multirow{2}[4]{*}{$h_1$} & \NCs\ $\als$ mod
 & \bf 114.70 & --- 
 & \it 89.76 & \it 88.83 & \bf 125.07 & \bf 126.71 & \it 117.71 & \it 117.71
 \\\cmidrule{2-10}
 & \NFHs\ $\alpha(\MZ)$
 & \bf 114.65 & ---
 & \it 89.72 & \it 89.31 & \bf 125.56 & \bf 125.74 & \it 118.31 & \it 117.77
 \\
 \midrule
 \midrule
 \multirow{2}[4]{*}{$h_2$} & \NCs\ $\als$ mod
 & \it 5951.36 & --- 
 & \bf 123.90 & \bf 125.87 & \it 142.96 & \it 142.74 & \bf 122.88 & \bf 123.60
 \\\cmidrule{2-10}
 & \NFHs\ $\alpha(\MZ)$
 & \it 5951.36 & --- 
 & \bf 124.26& \bf 124.88 & \it 142.98 & \it 142.59 & \bf 122.81 & \bf 123.08
 \\
 \midrule
 \midrule
 \multirow{2}[4]{*}{$h_3$} & \NCs\ $\als$ mod
 & 6370.76 & --- & 652.56 & 652.70 & 467.75 & 467.35 & 627.14 & 628.72
 \\\cmidrule{2-10}
 & \NFHs\ $\alpha(\MZ)$
 & 6370.90 & --- & 652.29 & 652.81 & 467.43 & 467.61 & 626.73 & 628.84
 \\\bottomrule
 \end{tabular}
 \caption{Mass predictions  for the  \cp-even scalars for  TP2--5 when
   using   modified    versions   of    \NC\   in   the    OS   option
   \NCs\ \OS\ (denoted shortly by \NCs) and \NFH\ (\NFHs).  Both codes
   are modified  to use an  identical numerical value for  $\als$.  In
   \NFH\      the       reparametrization      to      $\alpha{(\MZ)}$
   (\NFHs\  $\alpha{(\MZ)}$) is  used.  The values  correspond to  the
   two-loop result obtained with the OS renormalization scheme for the
   top/stop-sectors.  The  mass  values  for the  SM-like  scalar  are
   written  in  bold  fonts,  those for  the  singlet-like  scalar  in
   italics.  }
 \label{tab:resASmod}
\end{table}

\begin{table}[htb!]
\scriptsize \centering
\begin{tabular}{lllcccccc} 
 & & & \multicolumn{3}{c}{TP2} & \multicolumn{3}{c}{TP3}
 \\\toprule
 $i$ & $Q$ &
 & \ZHabs{i1} & \ZHabs{i2} & \ZHabs{i3}
 & \ZHabs{i1} & \ZHabs{i2} & \ZHabs{i3}
 \\\midrule\midrule
 \multirow{4}[4]{*}{$1$} & \multirow{2}[0]{*}{$\MS$}
 & \NCs\ $\als$ mod
 & {0.1034} & {0.9946} & {0.0004}
 & {0.2243} & {0.2140} & {0.9507} 
 \\ 
 & & \NFHs\ $\alpha(\MZ)$
 & 0.1034 & 0.9946 & 0.0004
 & 0.2229 & 0.2270 & 0.9480
 \\\cmidrule{2-9}
 & \multirow{2}[0]{*}{$\mt$}
 & \NCs\ $\als$ mod
 & {---} & {---} & {---}
 & {0.2234} & {0.2204} & {0.9495} 
 \\ 
 & & \NFHs\ $\alpha(\MZ)$
 & {---} & {---} & {---}
 & 0.2342 & 0.2610 & 0.9365
 \\\midrule\midrule
 \multirow{4}[4]{*}{$2$} & \multirow{2}[0]{*}{$\MS$}
 & \NCs\ $\als$ mod
 & {0.0096} & {0.0006} & {1.0000}
 & {0.2756} & {0.9218} & {0.2726}
 \\ 
 & & \NFHs\ $\alpha(\MZ)$
 & 0.0092 & 0.0005 & 1.0000
 & 0.2751 & 0.9183 & 0.2846
 \\\cmidrule{2-9}
 & \multirow{2}[0]{*}{$\mt$}
 & \NCs\ $\als$ mod
 & {---} & {---} & {---}
 & {0.2660} & {0.9233} & {0.2769} 
 \\
 & & \NFHs\ $\alpha(\MZ)$
 & {---} & {---} & {---}
 & 0.2566 & 0.9125 & 0.3185
 \\\midrule\midrule
 \multirow{4}[4]{*}{$3$} & \multirow{2}[0]{*}{$\MS$}
 & \NCs\ $\als$ mod
 & {0.9946} & {0.1034} & {0.0096}
 & {0.9347} & {0.3232} & {0.1478} 
 \\ 
 & & \NFHs\ $\alpha(\MZ)$
 & 0.9946 & 0.1034 & 0.0092
 & 0.9352 & 0.3243 & 0.1423
 \\\cmidrule{2-9}
 & \multirow{2}[0]{*}{$\mt$}
 & \NCs\ $\als$ mod
 & {---} & {---} & {---}
 & {0.9377} & {0.3144} & {0.1476} 
 \\ 
 & & \NFHs\ $\alpha(\MZ)$
 & {---} & {---} & {---}
 & 0.9377 & 0.3149 & 0.1467
 \\\bottomrule\bottomrule\\
\end{tabular}
\begin{tabular}{lllcccccc} 
 & & & \multicolumn{3}{c}{TP4} & \multicolumn{3}{c}{TP5}
 \\\toprule
 $i$ & $Q$ &
 & \ZHabs{i1} & \ZHabs{i2} & \ZHabs{i3}
 & \ZHabs{i1} & \ZHabs{i2} & \ZHabs{i3}
 \\\midrule\midrule
 \multirow{4}[4]{*}{$1$} & \multirow{2}[0]{*}{$\MS$}
 & \NCs\ $\als$ mod
 & {0.4839} & {0.7647} & {0.4255}
 & {0.1417} & {0.7305} & {0.6680} 
 \\ 
 & & \NFHs\ $\alpha(\MZ)$
 & 0.4872 & 0.8045 & 0.3398
 & 0.3319 & 0.6217 & 0.7094
 \\\cmidrule{2-9}
 & \multirow{2}[0]{*}{$\mt$}
 & \NCs\ $\als$ mod
 & {0.4766} & {0.7884} & {0.3900}
 & {0.3382} & {0.6914} & {0.6384} 
 \\ 
 & & \NFHs\ $\alpha(\MZ)$
 & 0.4779 & 0.8321 & 0.2813
 & 0.3476 & 0.7708 & 0.5339
 \\\midrule\midrule
 \multirow{4}[4]{*}{$2$} & \multirow{2}[0]{*}{$\MS$}
 & \NCs\ $\als$ mod
 & {0.0672} & {0.4524} & {0.8893}
 & {0.3315} & {0.6008} & {0.7274}
 \\ 
 & & \NFHs\ $\alpha(\MZ)$
 & 0.0270 & 0.3751 & 0.9266
 & 0.1379 & 0.7120 & 0.6885
 \\\cmidrule{2-9}
 & \multirow{2}[0]{*}{$\mt$}
 & \NCs\ $\als$ mod
 & {0.0413} & {0.4219} & {0.9057}
 & {0.0922} & {0.6508} & {0.7536} 
 \\
 & & \NFHs\ $\alpha(\MZ)$
 & 0.0122 & 0.3266 & 0.9451
 & 0.0482 & 0.5540 & 0.8311
 \\\midrule\midrule
 \multirow{4}[4]{*}{$3$} & \multirow{2}[0]{*}{$\MS$}
 & \NCs\ $\als$ mod
 & {0.8726} & {0.4589} & {0.1675}
 & {0.9327} & {0.3245} & {0.1570} 
 \\ 
 & & \NFHs\ $\alpha(\MZ)$
 & 0.8729 & 0.4606 & 0.1611
 & 0.9332 & 0.3264 & 0.1506
 \\\cmidrule{2-9}
 & \multirow{2}[0]{*}{$\mt$}
 & \NCs\ $\als$ mod
 & {0.8782} & {0.4477} & {0.1685}
 & {0.9365} & {0.3138} & {0.1564} 
 \\ 
 & & \NFHs\ $\alpha(\MZ)$
 & 0.8783 & 0.4482 & 0.1662
 & 0.9364 & 0.3147 & 0.1554
 \\\bottomrule
 \end{tabular}
 \label{tab:Mixing2L}
\caption{Two-loop  mixing matrix  elements  for TP2--5  when using  the
  indicated versions of \NC\ and \NFH\ as specified in the text, see \refse{sec:AsTreatment}.}
\end{table}


\subsection{Treatment of the strong coupling constant \boldmath{$\als$}}
\label{sec:AsTreatment}

\begin{figure}[htb!]
 \centering
 \includegraphics[width=.4\textwidth,width=.33\textheight]{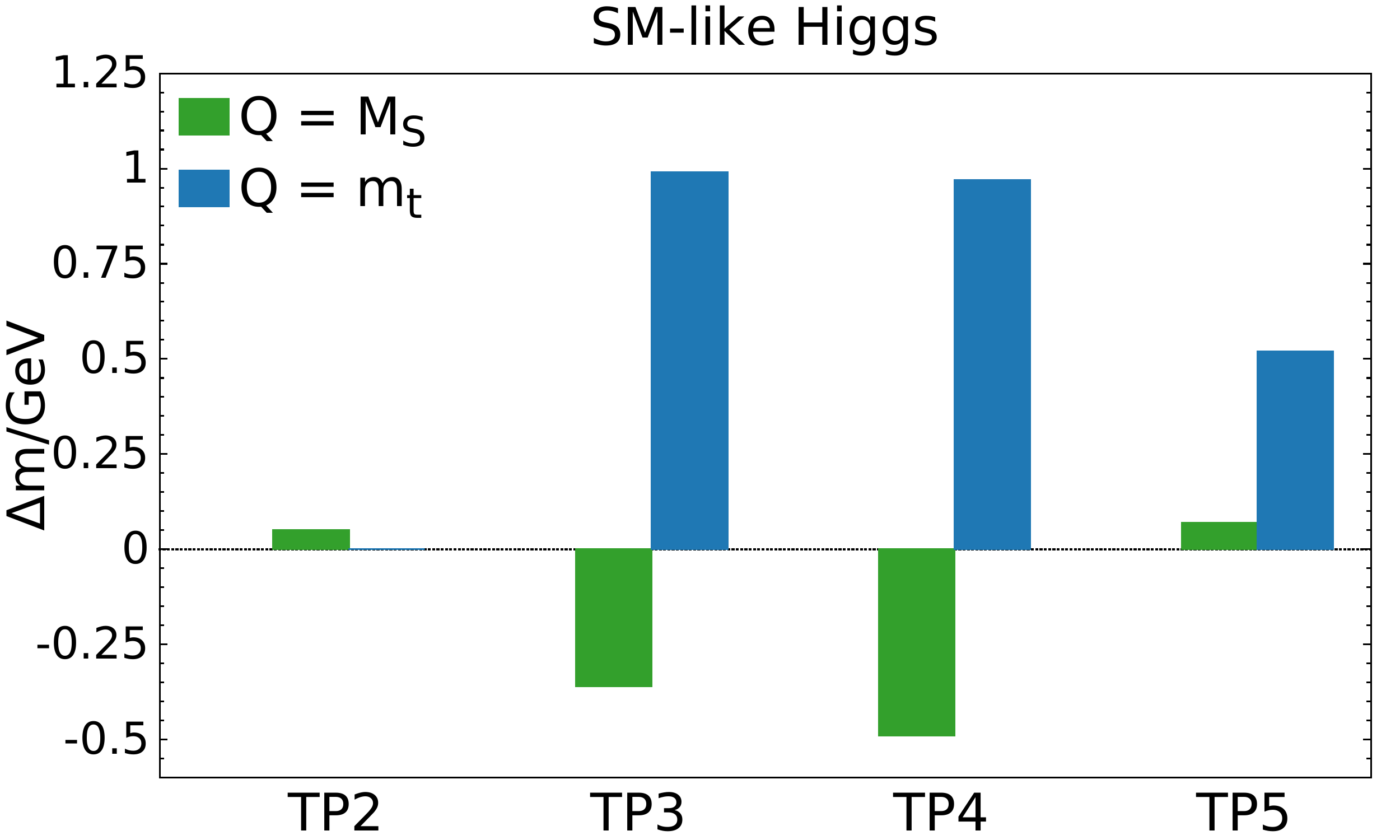}
 \;\;
 \includegraphics[width=.4\textwidth,width=.33\textheight]{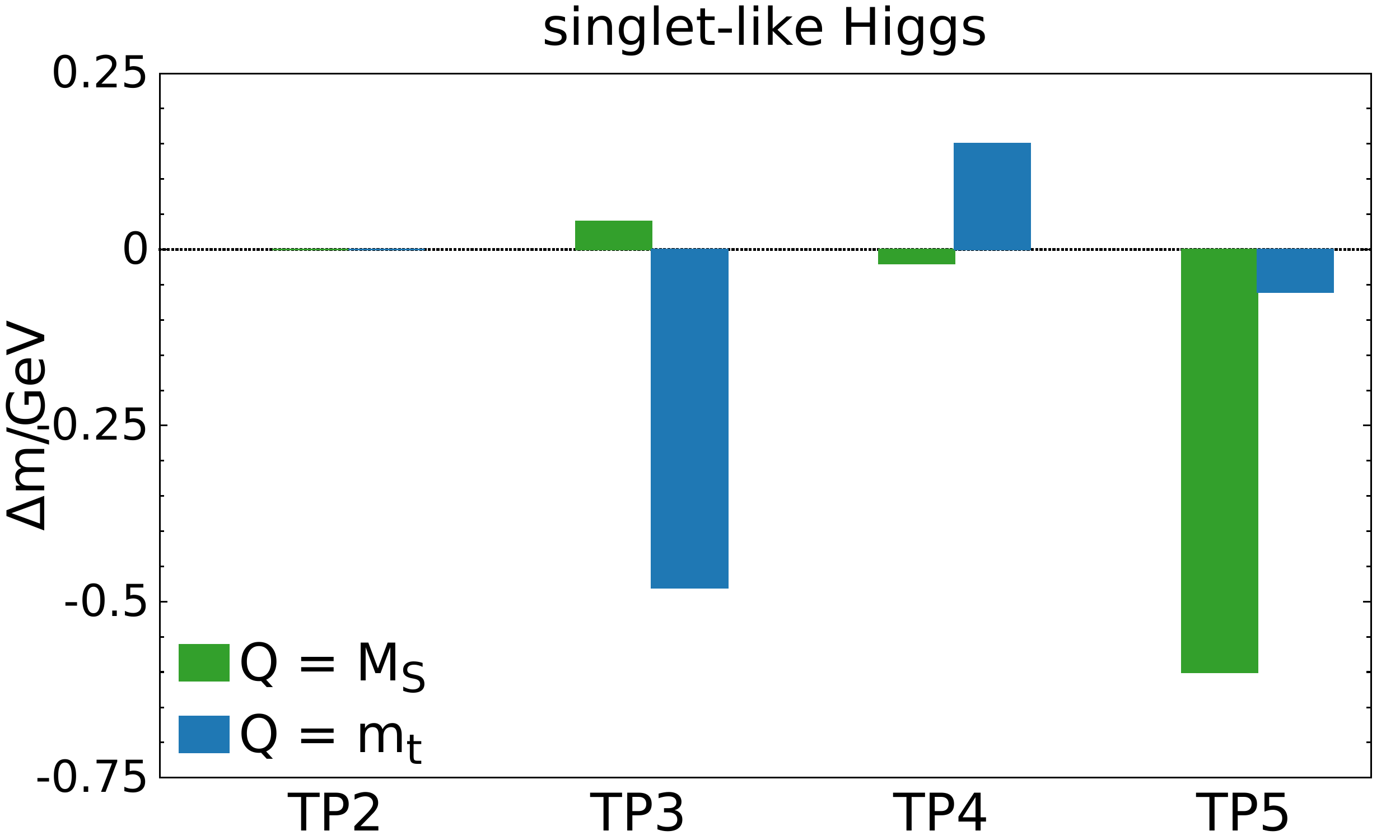}
 \\
 \includegraphics[width=.4\textwidth,width=.33\textheight]{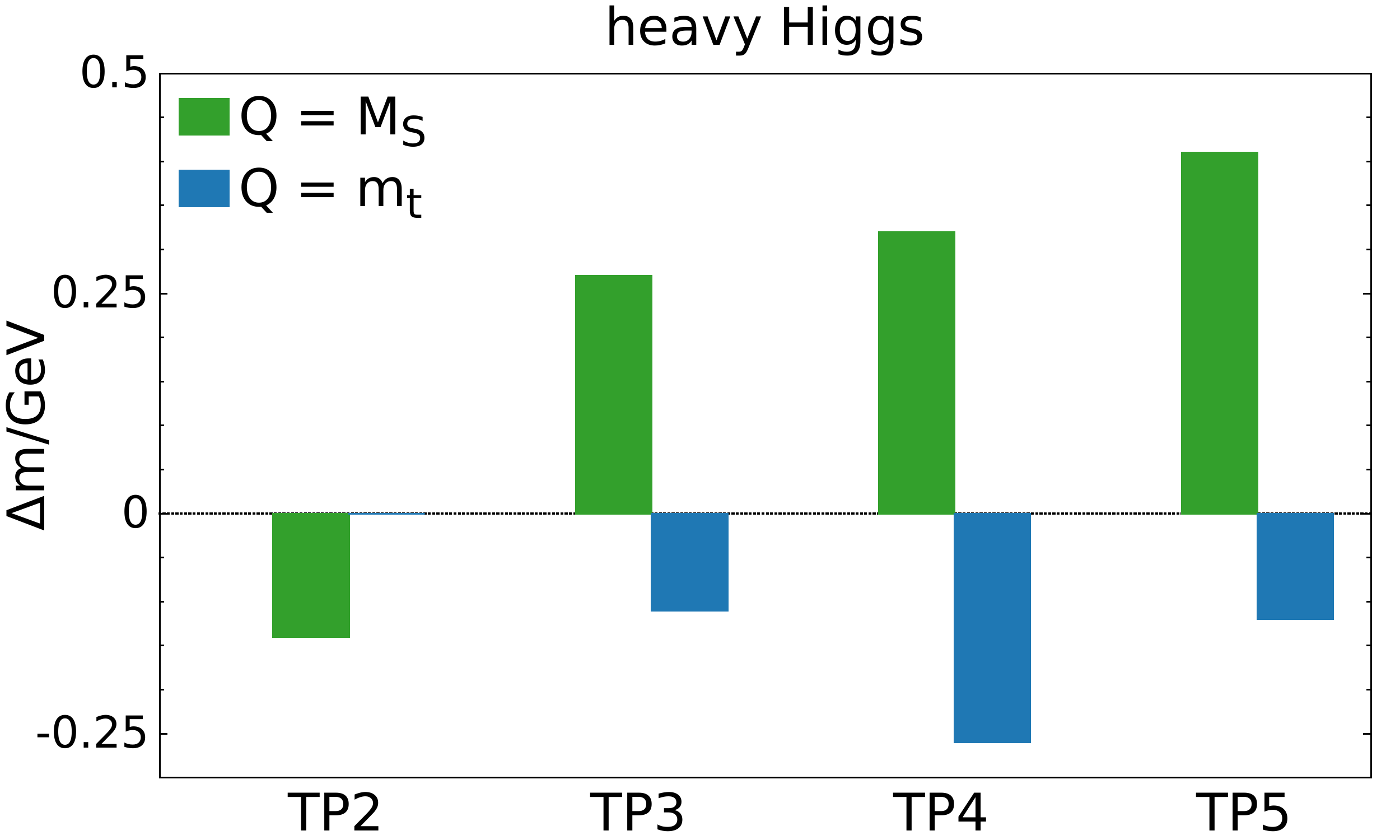}
 \caption{Difference $\Delta{m}  = M_h^{\NCs\ \alpha_s\  \text{mod}} -
   M_h^{\NFHs}$ between the Higgs  masses of the SM-like, singlet-like
   and  heavy Higgs  at the  two-loop level  calculated at  the scales
   \MS\ and \MTO.}
 \label{fig:asMod}
\end{figure}

In the following we analyze the  effects of the different treatment of
$\als$ in the two codes.  In \refta{tab:resASmod} the mass predictions
at \order{\alt\als} are given for the  case that \NC\ is modified such
that always the hard-coded \MSbar-value  of the strong coupling at the
scale  \MTO, $\als^{\MSbar}{(\MTO)}  = 0.10697$  as obtained  with the
routines of~\cite{Chetyrkin:2000yt}, is used. The results of \NFH\ are
not affected  by this procedure, since  $\als^{\MSbar}{(\MTO)}$ is the
standard value that is used. The  modified version of \NC\ is labelled
``\NCs\  $\als$  mod'' in  the  following.   As discussed  above,  for
\NFH\ we  continue to  use the  version reparametrized  to $\al(\MZ)$.
The  graphical  representation  of  the  differences  is  depicted  in
\reffi{fig:asMod}.  When  both codes use  the same numerical  value of
$\als$  a  better agreement  with  much  smaller relative  differences
between  their  mass predictions  can  be  observed.  The  differences
between the  mass predictions never  exceed 1.1~\gev, and  mostly stay
below  $0.5~\gev$  for  all  masses   in  all  TP-scenarios.   In  the
MSSM-limit we found again complete  agreement between the two codes at
the expected level of numerical accuracy now at \order{\alt\als}.

We conclude  that the main  source of  the observed difference  is the
different  treatment  of  the  strong coupling  constant  $\als$  (see
\refta{tab:DiffNcNfhVanilla}).     Although     the    renormalization
prescription and scale dependence of $\als$ represent effects formally
of three-loop order,  their effects on the Higgs  mass predictions can
be sizeable.   The corresponding mixing  matrix elements are  given in
\refta{tab:Mixing2L}.  In  the scenarios  TP2--4 at both  scales $\MS$
and $\mt$ we found differences of less than 5\% for the largest matrix
elements, which  contain the dominant  admixture to the  fields $h_i$.
Larger differences occur for the  subleading matrix elements, e.g.\ in
scenario TP3  differences of  up to  19\% can  be observed.   For TP4,
subleading matrix elements can differ even by up to a factor of 3. The
matrix elements  that differ  so strongly between  the two  codes are,
however, only a few percent of the  largest one. Even a change of them
by a factor of 3 results in relatively small differences compared with
the absolute  size of the  leading matrix elements.  For  the scenario
TP5 we  observe larger  differences for  the largest  matrix elements.
They differ by up to 20\%.  The  reason why the matrix elements of TP5
show larger discrepancies is the large mixing between the two lightest
\CP-even Higgs bosons where small  changes in the parameters can cause
a large effect  in the resulting mixing elements.  At  the scale $M_S$
the  adapted codes  show a  better  agreement for  the largest  mixing
matrix  elements when  compared to  their ``out-of-the-box''  versions
(cf.  tab.~\ref{tab:originalZH13_A}), where we found differences of up
to $\approx 40\%$.  At the scale  $\mt$ the adapted codes show a worse
agreement  with differences  of  up  to $\approx  20\%$  even for  the
largest matrix  elements, while  the corresponding difference  for the
``out-of-the-box'' versions never exceeded  3\%.  For scenario TP5 and
each of  the adapted codes,  however, we  find that the  mixing matrix
elements can differ  by up to $\approx 20\%$ when  evaluated at either
the scale  $M_S$ or $\mt$.  We  conclude that in the  scenario TP5 the
mixing matrix elements  are very sensitive to small  variations of the
parameters due  to the  large mixing between  the singlet  and SM-like
Higgs bosons.  The  results for the masses are much  less sensitive to
these effects.


\subsection{MSSM-Approximation beyond one-loop in \NFH}

In \NFH\ the  NMSSM contributions beyond one-loop  are approximated by
the respective corrections from the  MSSM at present.  This means that
at  \order{\alt\als}   the  genuine   NMSSM  contributions   are  only
incorporated in \NC,  as will be discussed below.  On  the other hand,
\NFH\    incorporates   further    MSSM-type   contributions    beyond
\order{\alt\als}. These  contributions consist of further  leading and
subleading         two-loop         corrections~\cite{Brignole:2001jy,
  Brignole:2002bz,  Degrassi:2002fi,   Dedes:2003km,  Borowka:2014wla,
  Borowka:2015ura} as well  as the resummation of  large logarithms to
all orders for high SUSY mass scales~\cite{Hahn:2013ria,Bahl:2016brp}.
In the MSSM  limit it has been found that  these corrections can yield
\order{5\,    \gev}   corrections    in    the   OS    renormalization
\cite{Brignole:2001jy, Brignole:2002bz, Degrassi:2002fi, Dedes:2003km,
  Borowka:2014wla, Borowka:2015ura, Hahn:2013ria, Bahl:2016brp}. This,
however, does not  take into account the impact of  non-zero values of
$\lambda$ which have not been evaluated in an OS calculation so far. A
\DRbar\   calculation  of   the  MSSM-approximated   $\order{(\alpha_t
  +\alpha_b)^2}$ contributions in \cite{Staub:2015aea} for TP1--5 gave
rise  to   a  $\sim  1\,\gev$  correction   (where  the  corresponding
$\order{\alpha_t  \alpha_s}$   calculation  yields   somewhat  smaller
corrections than our OS result), while the genuine NMSSM contributions
from   the  fermion/sfermion-   and   Higgs/Higgsino-sectors  in   the
electroweak gauge-less  limit \cite{Goodsell:2014pla} gave rise  to an
additional  $\lesssim 1\,\gev$  correction. We  leave a  more detailed
discussion for future work.

\begin{figure}[htb!]
 \centering
 \includegraphics[width=.4\textwidth,width=.33\textheight]{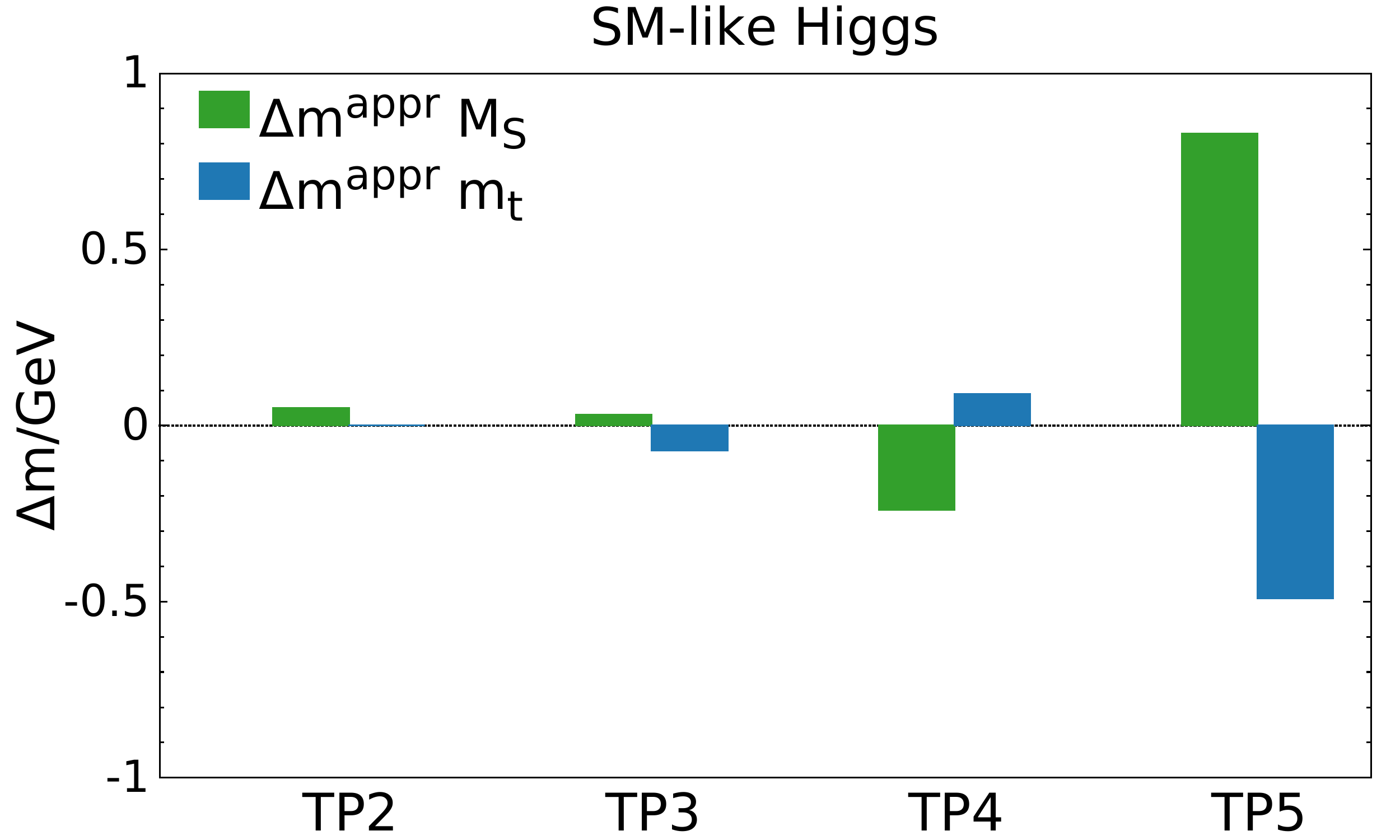}
 \;\;
 \includegraphics[width=.4\textwidth,width=.33\textheight]{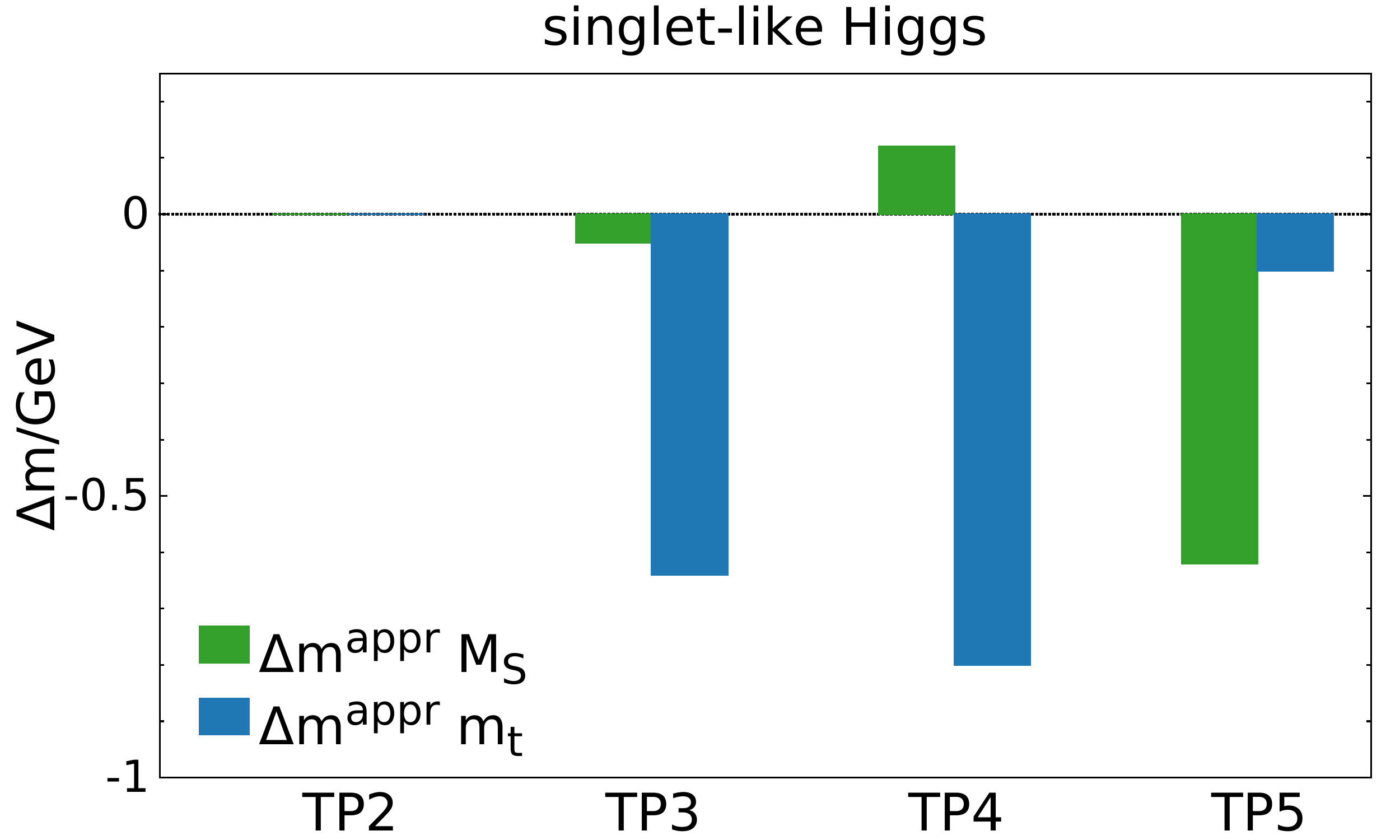}
 \;\;
 \\
 \includegraphics[width=.4\textwidth,width=.33\textheight]{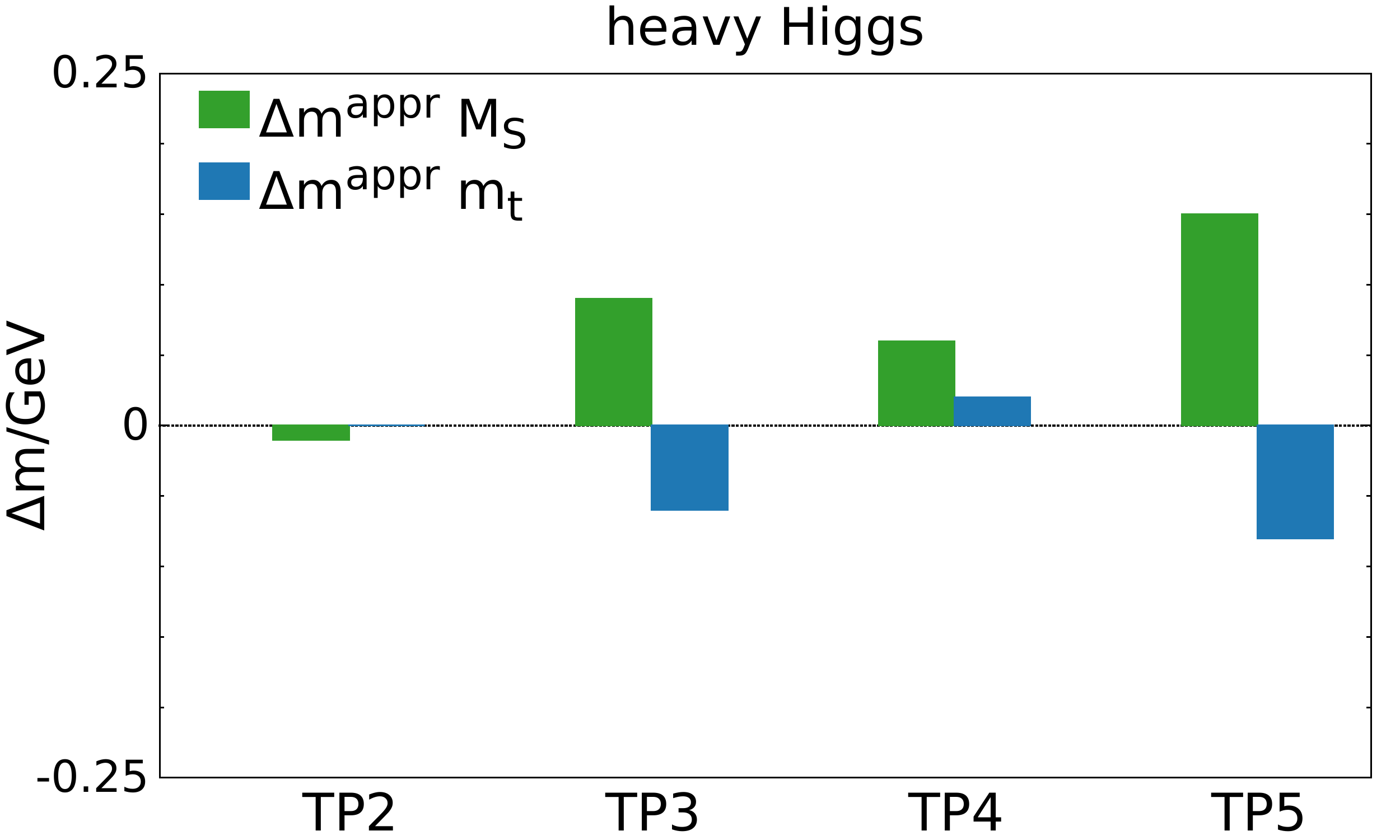}
 \\
 \caption{$\Delta{m}^{\rm appr} =  \Delta{m}^{\rm 2L} - \Delta{m}^{\rm
     1L}$: size  of the effect  of the MSSM-approximation  employed at
   \order{\alt\als} in  \NFH. The  used version  of \NFH\  employs the
   reparametrization to $\alpha{(\MZ)}$. }
 \label{fig:ApproxMSSM}
\end{figure}

At   \order{\alt\als}   the   genuine   NMSSM   two-loop   corrections
incorporated in  \NC\ give rise to  differences to \NFH.  In  order to
estimate their impact we compare the two-loop mass predictions between
'\NCs\   $\als$   mod'   and   '\NFHs\   $\alpha{(\MZ)}$'   given   in
\refta{tab:resASmod}.    The   effect    of   the   MSSM-approximation
$\Delta{m}^{\rm appr}$ can be obtained by
\begin{align}
 \Delta{m}^{\rm appr} = \Delta{m}^{\rm 2L} - \Delta{m}^{\rm 1L},
\end{align}
\noindent
where  the  $\Delta{m}^{n{\rm L}}$  are  the  differences between  the
result of  \NC\ and \NFH\ at  the $n$-th loop order  obtained from the
results   given  in   tabs.~\ref{tab:resOB1L}  (where   we  take   the
'\NFHs\  $\alpha{(\MZ)}$' value  for \NFH)  and~\ref{tab:resASmod}. By
this construction the effects of the residual differences arising from
the  different  treatment  of the  electromagnetic  coupling  constant
$\alpha$  are separated  from the  effects of  the MSSM-approximation.
The     results    of     these    comparisons     are    shown     in
\reffi{fig:ApproxMSSM}. As expected the  approximation has the largest
effects for the scenarios TP3--5  with large values of $\lambda$.  For
the SM-like Higgs-field $\Delta{m}_h^{\rm  appr}$ does not exceed $\pm
750 (500)~\mev$ at  the scale $\MS (\mt)$, shown as  dark green (blue)
bars.  For the singlet-like Higgs-boson it stays below $\pm 750\,\mev$
for both  scales, and for  the heavy  Higgs field we  find differences
below $\pm 200\,\mev$.  This is in accordance with the expected impact
of the  approximation as described in  ref.~\cite{Drechsel:2016jdg} as
well as with the results of ref.~\cite{Muhlleitner:2014vsa}.


%% file: sections/calculationNC.tex
\subsection{Comparison with \boldmath{\DRbar} calculation in \NC}\label{sec:DRbar}

As  a final  step  we now  compare  between different  renormalization
schemes.  For  all our  results shown  up to  now we  have used  an OS
renormalization   of   the   parameters  in   the   top/stop   sector.
\NC\ offers,  however, also the  possibility to switch between  OS and
\DRbar\  renormalization of  the  top/stop sector,  which affects  the
\order{\alt\als} corrections.   In this  section the default  value of
\NC\ for $\alpha_s$ in the \DRbar\ scheme at the scale $Q$ is used.

\begin{table}[h]
  \scriptsize \centering
  \begin{tabular}{llcccccccc}
    &
    & \multicolumn{2}{c}{TP2} & \multicolumn{2}{c}{TP3}
    & \multicolumn{2}{c}{TP4} & \multicolumn{2}{c}{TP5}
    \\\toprule
    & $Q$ & $\MS$ & $\mt$ & $\MS$ & $\mt$
    & $\MS$ & $\mt$ & $\MS$ & $\mt$
    \\\midrule\midrule
    \multirow{2}[4]{*}{$h_1$} 
    & \NCs\ \OS
    & \bf 120.42 & --- 
    & \it 89.92 & \it 88.81 & \bf 126.44 & \bf 126.65 & \it 119.54 & \it 117.63
    \\\cmidrule{2-10}
    & \NCs\ \DRbar
     & \bf 118.57 & ---
    & \it 90.17 & \it 88.94 & \bf 126.15 & \bf 125.90 & \it 119.86 & \it 118.56
    \\
    \midrule
    \midrule
    \multirow{2}[4]{*}{$h_2$} 
    & \NCs\ \OS
   & \it 5951.36 & --- 
    & \bf 126.16 & \bf 125.80 & \it 143.32 & \it 142.73 & \bf 124.44 & \bf 123.51
    \\\cmidrule{2-10}
    & \NCs\ \DRbar
    & \it 5951.36 & --- 
    & \bf 124.88 & \bf 124.86 & \it 142.38 & \it 142.16 & \bf 123.28 & \bf 125.20
    \\
    \midrule
    \midrule
    \multirow{2}[4]{*}{$h_3$} 
    & \NCs\ \OS
     & 6370.77 & --- & 652.60 & 652.70 & 467.89 & 467.35 & 627.18 & 628.72
    \\\cmidrule{2-10}
    & \NCs\ \DRbar
     & 6371.21 & --- & 652.44 & 652.65 & 467.39 & 467.03 & 626.97 & 628.75
    \\\bottomrule
  \end{tabular}
 \caption{Mass predictions  for the  \cp-even scalars for  TP2--5 when
   using the  OS and \DRbar\  renormalization in the  top/stop sector.
   The mass  values of the SM-like  scalar are written in  bold fonts,
   those of the singlet-like scalar in italics.}
 \label{tab:resDRbar}
\end{table}
%
\begin{figure}[h]
 \centering
 \includegraphics[width=.4\textwidth,width=.33\textheight]{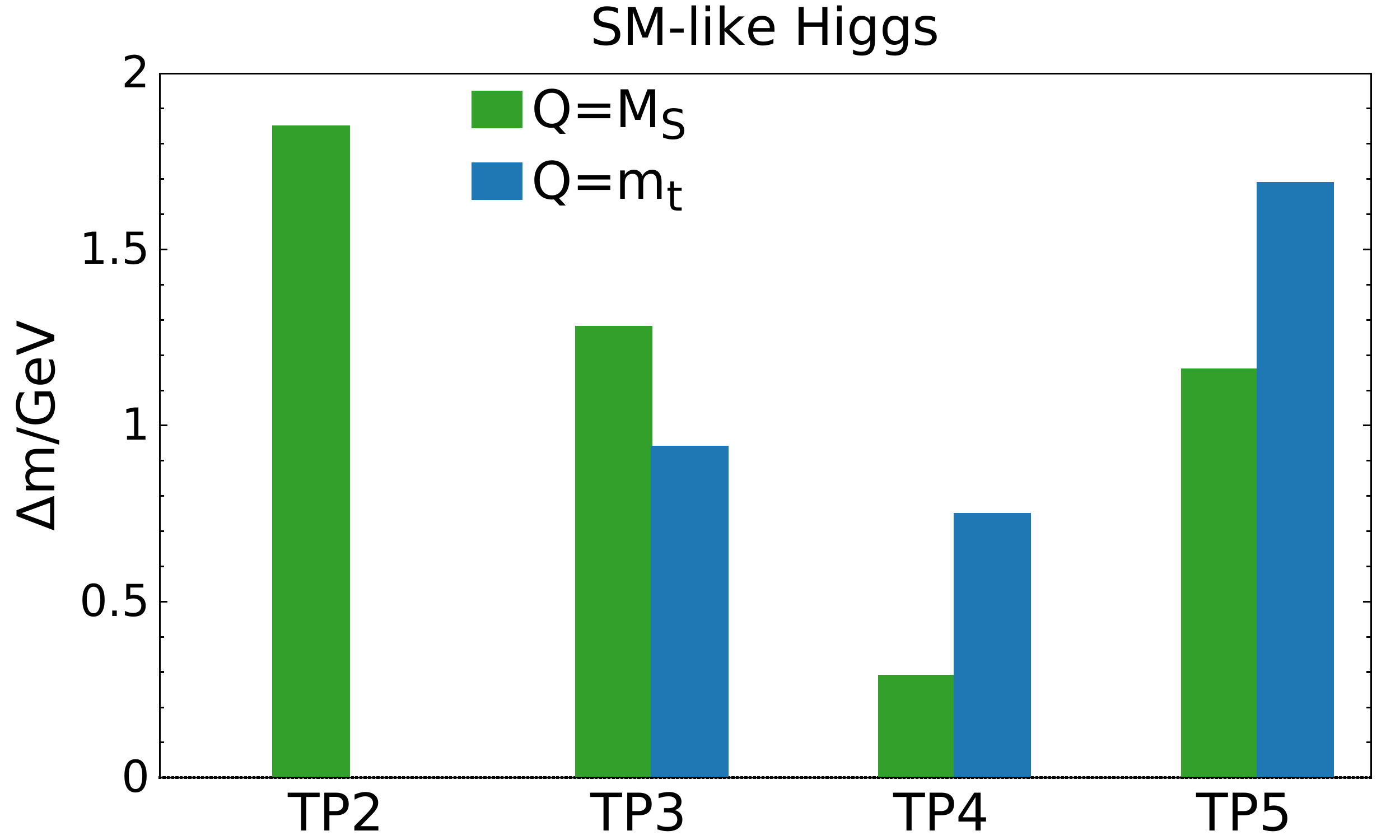}
 \;\;
 \includegraphics[width=.4\textwidth,width=.33\textheight]{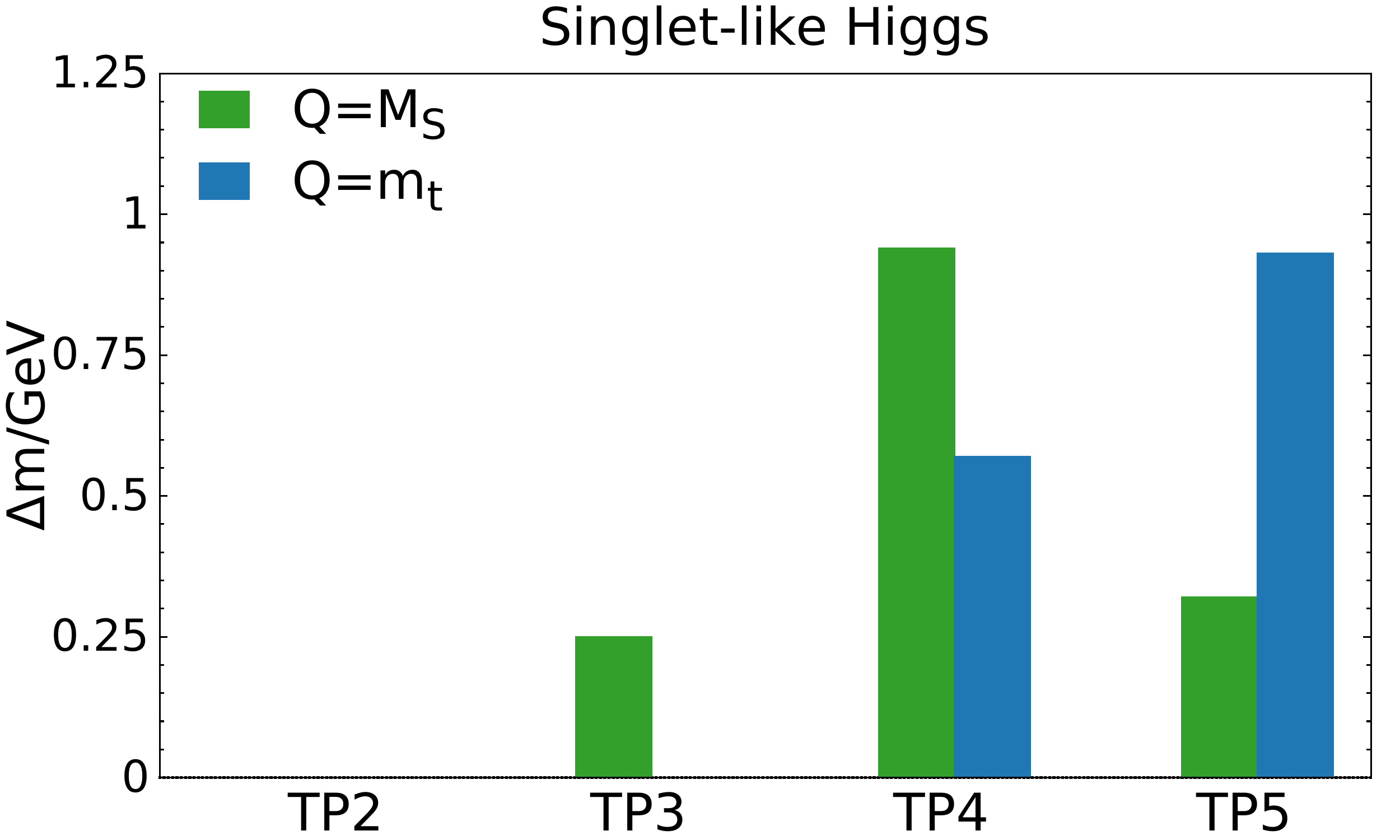}
 \\
 \includegraphics[width=.4\textwidth,width=.33\textheight]{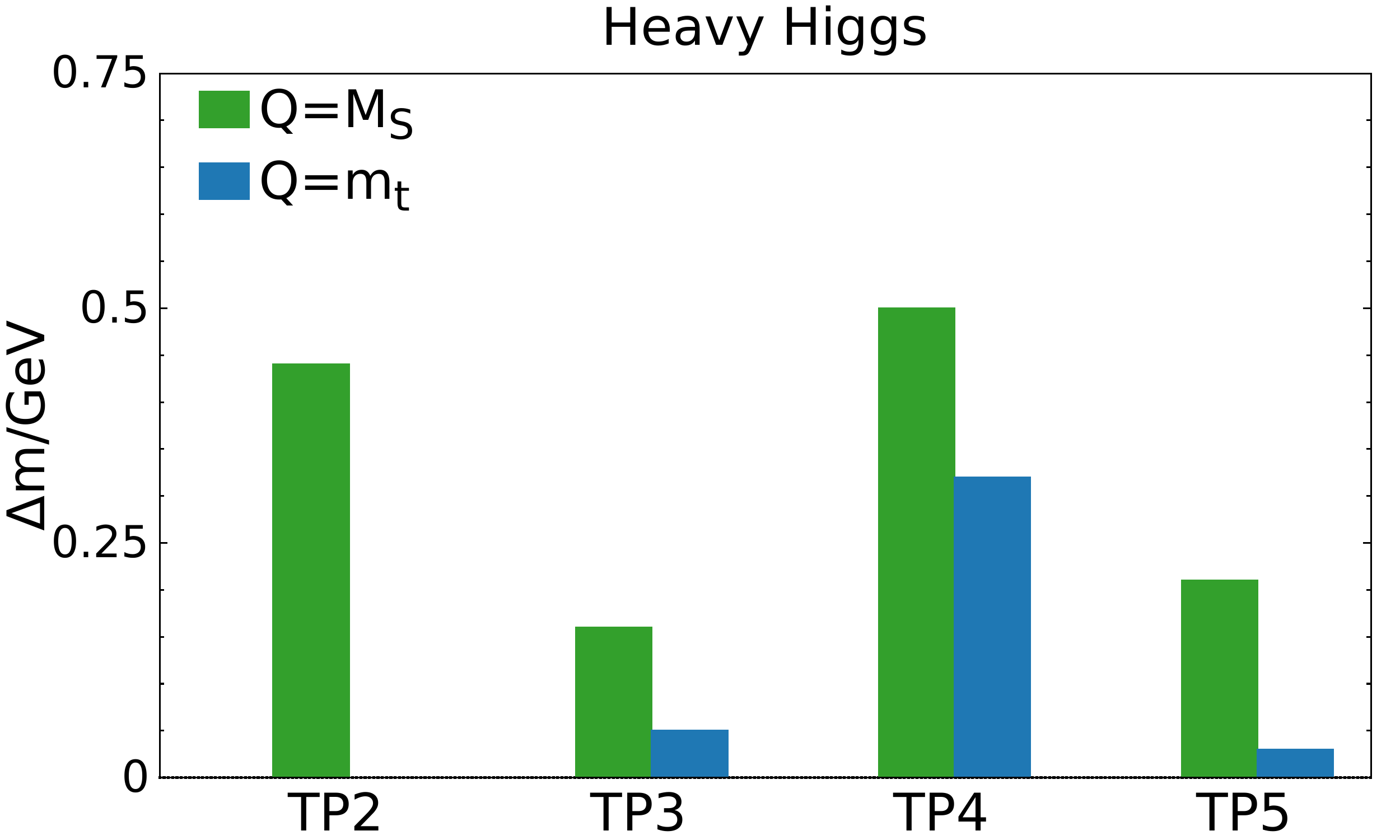}
 \caption{Difference   $\Delta{m}    =   M_h^{\NCs\   {\rm    OS}}   -
   M_h^{\NCs\   \DRbar}$   from  tab.~\ref{tab:resDRbar}.    For   TP2
   higher-order corrections to the mass  of the singlet-like field are
   suppressed due to the small value of $\lambda$.}
 \label{fig:total}
\end{figure}
\begin{table}[h]
 \scriptsize
 \centering
 \begin{tabular}{crrrrrrrr}
 & \multicolumn{2}{c}{TP2} & \multicolumn{2}{c}{TP3}
 & \multicolumn{2}{c}{TP4} & \multicolumn{2}{c}{TP5}
 \\\toprule
 $Q$ & $\MS$ & $\mt$ & $\MS$ & $\mt$
 & $\MS$ & $\mt$ & $\MS$ & $\mt$
 \\\midrule\midrule
$m_t^{\DRbar}$ & 136.7 & --- & 143.9 & 154.2 & 146.9 & 155.8 & 140.0 &  152.9
\\ \midrule
 $m_{\tilde{t}_1}^{\rm \DRbar}$ & $488.4$ & --- & $940.8$ & $1195$ & $671.8$ & $839.1$
 & $1503$ & $1548$
 \\\midrule
 $m_{\tilde{t}_2}^{\rm \DRbar}$ & $2509$ & --- & $1074$ & $1267$ & $845.3$ & $949.7$
 & $1509$ & $1591$
 \\\bottomrule
 \end{tabular}
 \caption{\DRbar\ top and stop masses, given in GeV, in the
   TP2--5 scenarios obtained by the routines of \NC. 
}
 \label{tab:DRbarstopmasses}
\end{table}

\begin{table}[htb!]
  \scriptsize \centering
  \begin{tabular}{lllcccccc}
    & & & \multicolumn{3}{c}{TP2} & \multicolumn{3}{c}{TP3}
    \\\toprule
    $i$ & $Q$ &
    & \ZHabs{i1} & \ZHabs{i2} &  \ZHabs{i3}
    & \ZHabs{i1} & \ZHabs{i2} &  \ZHabs{i3}
    \\\midrule\midrule
    \multirow{4}[4]{*}{$1$} & \multirow{2}[0]{*}{$ \MS$}
    & \NCs\ \OS  
    & {0.1034} & {0.9946} & {0.0004}
    & {0.2199} & {0.1994} & {0.9549}
    \\ 
    & & \NCs\ \DRbar\
    & 0.1034 & 0.9946 & 0.0004
    & 0.2189 & 0.1962 & 0.9558
    \\\cmidrule{2-9}
    & \multirow{2}[0]{*}{$\mt$}
    & \NCs\ \OS
    & {---} & {---} & {---} 
    & {0.2236} & {0.2210} & {0.9493}
    \\ 
    & & \NCs\ \DRbar\ 
    & {---} & {---} & {---}
    & 0.2237 & 0.2221 & 0.9490
    \\\midrule\midrule
    \multirow{4}[4]{*}{$2$} & \multirow{2}[0]{*}{$ \MS$}
    & \NCs\ \OS
    & {0.0096} & {0.0006} & {1.}
    & {0.2797} & {0.9249} & {0.2575}
    \\&&  \NCs\ \DRbar\
    & 0.0095 & 0.0006 & 1.
    & 0.2802 & 0.9257 & 0.2542
    \\\cmidrule{2-9}
    & \multirow{2}[0]{*}{$\mt$}
    & \NCs\ \OS
    & {---} & {---} & {---}
    & {0.2659} & {0.9232} & {0.2775}
    \\
    & & \NCs\ \DRbar\
    & {---} & {---} & {---}
    & 0.2656 & 0.9230 & 0.2786
    \\\midrule\midrule
    \multirow{4}[4]{*}{$3$} & \multirow{2}[0]{*}{$\MS$}
    & \NCs\ \OS
    & {0.9946} & {0.1034} & {0.0096}
    & {0.9346} & {0.3237} & {0.1476}
    \\ 
    & & \NCs\ \DRbar\
    & 0.9946 & 0.1034 & 0.0095
    & 0.9346 & 0.3235 & 0.1476
    \\\cmidrule{2-9}
    & \multirow{2}[0]{*}{$\mt$}
    & \NCs\ \OS
    & {---} & {---} & {---}
    & {0.9377} & {0.3144} & {0.1476}
    \\
    & & \NCs\ \DRbar\
    & {---} & {---} & {---}
   & 0.9378 & 0.3144 & 0.1475
    \\\bottomrule\bottomrule\\
  \end{tabular}
  \\
 \begin{tabular}{lllcccccc} 
    & & & \multicolumn{3}{c}{TP4} & \multicolumn{3}{c}{TP5}
    \\\toprule
    $i$ & $Q$ &
    & \ZHabs{i1} & \ZHabs{i2} &  \ZHabs{i3}
    & \ZHabs{i1} & \ZHabs{i2} &  \ZHabs{i3}
    \\\midrule\midrule
    \multirow{4}[4]{*}{$1$} & \multirow{2}[0]{*}{$\MS$}
    & \NCs\ \OS
    & {0.4813} & {0.7432} & {0.4648}
    & {0.2845} & {0.3943} & {0.8738}
    \\ 
    & & \NCs\ \DRbar\
    & 0.4863 & 0.7790 & 0.3959
    & 0.2994 & 0.4536 & 0.8394
    \\\cmidrule{2-9}
    & \multirow{2}[0]{*}{$\mt$}
    & \NCs\ \OS
    & {0.4766} & {0.7886} & {0.3885}
    & {0.3393} & {0.6991} & {0.6294}
    \\ 
    & & \NCs\ \DRbar\
    & 0.4777 & 0.8098 & 0.3407
    & 0.3114 & 0.5379 & 0.7834
    \\\midrule\midrule
    \multirow{4}[4]{*}{$2$} & \multirow{2}[0]{*}{$\MS$}
    & \NCs\ \OS
    & {0.0895} & {0.4858} & {0.8694}
    & {0.2224} & {0.8594} & {0.4603}
    \\ 
    & & \NCs\ \DRbar\
    & 0.0515 & 0.4267 & 0.9029
    & 0.2017 & 0.8298 & 0.5203
    \\\cmidrule{2-9}
    & \multirow{2}[0]{*}{$\mt$}
    & \NCs\ \OS
    & {0.0411} & {0.4215} & {0.9069}
    & {0.0882} & {0.6425} & {0.7612}
    \\
    & & \NCs\ \DRbar\
    & 0.0162 & 0.3796 & 0.9250
    & 0.1623 & 0.7822 & 0.6015
    \\\midrule\midrule
    \multirow{4}[4]{*}{$3$} & \multirow{2}[0]{*}{$\MS$}
    & \NCs\ \OS 
    & {0.8720} & {0.4600} & {0.1673}
    & {0.9325} & {0.3253} & {0.1568}
    \\ 
    & & \NCs\ \DRbar\
    & 0.8723 & 0.4595 & 0.1674
    & 0.9326 & 0.3251 & 0.1569
    \\\cmidrule{2-9}
    & \multirow{2}[0]{*}{$\mt$}
    & \NCs\ \OS
    & {0.8782} & {0.4477} & {0.1685}
    & {0.9365} & {0.3137} & {0.1564}
    \\ 
    & & \NCs\ \DRbar\
    & 0.8784 & 0.4474 & 0.1682
    & 0.9363 & 0.3144 & 0.1563
    \\\bottomrule
  \end{tabular}
 \caption{Absolute  values  for  the  mixing matrix  elements  of  the
 \cp-even scalar sector for TP2--5  
when using the OS or the \DRbar\ renormalization
of the top/stop sector within \NC.
  \label{tab:ZH13DRbar}
}
\end{table}

In tab.~\ref{tab:resDRbar} the predictions for the neutral Higgs boson
masses  from  \NC\  with  OS renormalization  (first  line)  and  with
\DRbar\ renormalization (second line) of the top/stop sector are given
for TP2--5.\footnote{Note  that wherever we  give values at  the scale
  $Q=m_t$ the numerical value of the scale is taken to be the top pole
  mass,  $m_t=  172.9$~GeV.}   The  numbers corresponding  to  the  OS
renormalization of the top/stop  sector in tab.~\ref{tab:resDRbar} are
identical   to  the   \NC\  results   in  tab.~\ref{tab:resOB}.    The
differences  in   the  Higgs   masses  due  to   the  change   of  the
renormalization  scheme  between  top/stop sector  are  visualized  in
fig.~\ref{fig:total}.  The  values of the  stop and top masses  in the
\DRbar\  scheme can  be found  in tab.~\ref{tab:DRbarstopmasses},  the
stop masses as obtained in the OS scheme in tab.~\ref{tab:OS}.

As    can    be     inferred    from    tab.~\ref{tab:resDRbar}    and
fig.~\ref{fig:total},  the different  renormalization schemes  lead in
general  to   differences  of  \order{1~\gev}  for   the  SM-like  and
singlet-like Higgs boson  with a maximum difference  of $1.9~\gev$ for
the  SM-like Higgs  boson.  For  the  heavy Higgs  bosons the  maximum
difference reaches up  to $0.5~\gev$. The effects are  of similar size
for  both  the scale  $Q  =  \MS$ and  $Q  =  \mt$ and  most  strongly
pronounced for the SM-like Higgs boson,  which is affected most by the
corrections  of the  top/stop sector  as it  has the  largest $\phi_2$
component,  the  component  that   couples  to  up-type  quarks.   The
numerical  differences between  the different  renormalization schemes
are indicative  of the  theoretical uncertainties  due to  the missing
higher order  corrections.  However, one  should keep in mind  that in
\refse{sec:AsTreatment} we found  that using a different  $\als$ has a
larger  impact on  the Higgs  boson mass  than the  difference due  to
different renormalization  schemes of the top/stop  sector.  Since the
scale choice of  $\als$ is formally a higher-order  effect this points
to a larger higher-order uncertainty than the one we obtain here.

In  tab.~\ref{tab:ZH13DRbar}  the  mixing   matrix  elements  for  the
different options of the renormalization of the top/stop sector can be
found.  For TP2 the renormalization  scheme has basically no influence
on  the  mixing  matrix  elements.   For  TP3  the  influence  of  the
renormalization scheme is  well below 2\%, whereas for TP4  and TP5 in
some cases  the renormalization  scheme can  change the  mixing matrix
elements by more than a factor  two.  These large differences occur in
the smallest mixing  matrix element of the respective  Higgs boson for
the singlet-like  Higgs as well as  for a SM-like Higgs  with sizeable
singlet admixture.  For most of the  matrix elements the change due to
the renormalization scheme is, however, well below 10\%.

Finally,  we  also  want  to  make  contact  with  the  discussion  in
ref.~\cite{Staub:2015aea}.  Contrary  to our scenarios, where  we used
$\MHp$  as  input, in  ref.~\cite{Staub:2015aea}  $\Ala$  was used  as
input. This corresponds to a slightly different renormalization scheme
in  \NC.  If $\MHp$  is  used  as input,  the  charged  Higgs mass  is
renormalized OS and subsequently $\Ala$ is determined from the charged
Higgs mass,  whereas if $\Ala$  is given  as input it  is renormalized
\DRbar.  In tab.~\ref{tab:resAlam}  we show values where  the input is
given by $\Ala$ (first line) and  by $\MHp$ (second line).  All values
in tab.~\ref{tab:resAlam}  are given  for the  \DRbar\ renormalization
scheme  of the  top/stop sector.   The first  line corresponds  to the
"out-of-the-box"  \NC\ values  as given  in ref.~\cite{Staub:2015aea}.
The effect of the way  $\Ala$ (or respectively $\MHp$) is renormalized
is small.   Only for the most  singlet-like Higgs boson it  can exceed
1~GeV.  For all the other Higgs  bosons it is always well below 1~GeV,
and in particular  for the SM-like Higgs  boson it is at  the level of
\order{100~\mev}.  In tab.~\ref{tab:ZH13Alam} the values of the mixing
matrix elements are given.  Like  for the Higgs masses the differences
between the  input $\Ala$ or  $\MHp$ is  small.  It should  finally be
noted that if $\MHp$  is input in \NC\, the $\Ala$  in the output file
is determined from $\MHp$ at tree  level. This implies that the $\Ala$
in the  output of the computation  with input $\MHp$ will  differ from
the $\Ala$ given in ref.~\cite{Staub:2015aea}.

\begin{table}[h]
  \scriptsize \centering
  \begin{tabular}{llcccccccc}
    &
    & \multicolumn{2}{c}{TP2} & \multicolumn{2}{c}{TP3}
    & \multicolumn{2}{c}{TP4} & \multicolumn{2}{c}{TP5}
    \\\toprule
    & $Q$ & $\MS$ & $\mt$ & $\MS$ & $\mt$
    & $\MS$ & $\mt$ & $\MS$ & $\mt$
    \\\midrule\midrule
    \multirow{2}[4]{*}{$h_1$} 
    & \NCs\ $\Ala$
    & \bf 118.57 & --- 
    & \it 90.88 & \it 87.78 & \bf 126.37 & \bf 125.76 & \it 120.32 & \it 118.65
    \\\cmidrule{2-10}
    & \NCs\ $\MHp$
     & \bf 118.57 & ---
    & \it 90.17 & \it 88.94 & \bf 126.15 & \bf 125.90 & \it 119.86 & \it 118.56
    \\
    \midrule
    \midrule
    \multirow{2}[4]{*}{$h_2$} 
& \NCs\ $\Ala$
   & \it 5951.36 & --- 
    & \bf 124.86 & \bf 124.68 & \it 142.59 & \it 141.28 & \bf 123.14 & \bf 125.26
    \\\cmidrule{2-10}
    & \NCs\ $\MHp$
    & \it 5951.36 & --- 
    & \bf 124.88 & \bf 124.86 & \it 142.38 & \it 142.16 & \bf 123.28 & \bf 125.20
    \\
    \midrule
    \midrule
    \multirow{2}[4]{*}{$h_3$} 
    & \NCs\ $\Ala$
     & 6371.31 & --- & 652.48 & 652.64 & 467.42 & 467.01 & 627.00 & 628.77
    \\\cmidrule{2-10}
    & \NCs\ $\MHp$
     & 6371.21 & --- & 652.44 & 652.65 & 467.39 & 467.03 & 626.97 & 628.75
    \\\bottomrule
  \end{tabular}
 \caption{Mass predictions  for the  \cp-even scalars for  TP2--5 when
   using the \DRbar\ renormalization in the top/stop sector for either
   $\Ala$ as input  or $\MHp$.  The mass values of  the SM-like scalar
   are  written in  bold fonts,  those of  the singlet-like  scalar in
   italics.}
  \label{tab:resAlam}
\end{table}
\begin{table}[htb!]
  \scriptsize \centering
  \begin{tabular}{lllcccccc}
    & & & \multicolumn{3}{c}{TP2} & \multicolumn{3}{c}{TP3}
    \\\toprule
    $i$ & $Q$ &
    & \ZHabs{i1} & \ZHabs{i2} &  \ZHabs{i3}
    & \ZHabs{i1} & \ZHabs{i2} &  \ZHabs{i3}
    \\\midrule\midrule
    \multirow{4}[4]{*}{$1$} & \multirow{2}[0]{*}{$ \MS$}
    & \NCs\ $\Ala$  
    & {0.1034} & {0.9946} & {0.0004}
    & {0.2177} & {0.1923} & {0.9569}
    \\ 
    & & \NCs\ $\MHp$
    & 0.1034 & 0.9946 & 0.0004
    & 0.2189 & 0.1962 & 0.9558
    \\\cmidrule{2-9}
    & \multirow{2}[0]{*}{$\mt$}
    & \NCs\ $\Ala$
    & {---} & {---} & {---} 
    & {0.2212} & {0.2142} & {0.9514}
    \\ 
    & & \NCs\ $\MHp$
    & {---} & {---} & {---}
    & 0.2237 & 0.2221 & 0.9490
    \\\midrule\midrule
    \multirow{4}[4]{*}{$2$} & \multirow{2}[0]{*}{$ \MS$}
    & \NCs\ $\Ala$
    & {0.0095} & {0.0006} & {1.}
    & {0.2811} & {0.9265} & {0.2502}
    \\&&  \NCs\ $\MHp$
    & 0.0095 & 0.0006 & 1.
    & 0.2802 & 0.9257 & 0.2542
    \\\cmidrule{2-9}
    & \multirow{2}[0]{*}{$\mt$}
    & \NCs\ $\Ala$
    & {---} & {---} & {---}
    & {0.2677} & {0.9248} & {0.2705}
    \\
    & & \NCs\ $\MHp$
    & {---} & {---} & {---}
    & 0.2656 & 0.9230 & 0.2786
    \\\midrule\midrule
    \multirow{4}[4]{*}{$3$} & \multirow{2}[0]{*}{$\MS$}
    & \NCs\ $\Ala$
    & {0.9946} & {0.1034} & {0.0095}
    & {0.9347} & {0.3234} & {0.1476}
    \\ 
    & & \NCs\ $\MHp$
    & 0.9946 & 0.1034 & 0.0095
    & 0.9346 & 0.3235 & 0.1476
    \\\cmidrule{2-9}
    & \multirow{2}[0]{*}{$\mt$}
    & \NCs\ $\Ala$
    & {---} & {---} & {---}
    & {0.9378} & {0.3145} & {0.1472}
    \\
    & & \NCs\ $\MHp$
    & {---} & {---} & {---}
    & 0.9378 & 0.3144 & 0.1475
    \\\bottomrule\bottomrule\\
  \end{tabular}
  \\
 \begin{tabular}{lllcccccc} 
    & & & \multicolumn{3}{c}{TP4} & \multicolumn{3}{c}{TP5}
    \\\toprule
    $i$ & $Q$ &
    & \ZHabs{i1} & \ZHabs{i2} &  \ZHabs{i3}
    & \ZHabs{i1} & \ZHabs{i2} &  \ZHabs{i3}
    \\\midrule\midrule
    \multirow{4}[4]{*}{$1$} & \multirow{2}[0]{*}{$\MS$}
    & \NCs\ $\Ala$
    & {0.4869} & {0.7849} & {0.3832}
    & {0.2967} & {0.4433} & {0.8459}
    \\ 
    & & \NCs\ $\MHp$
    & 0.4863 & 0.7790 & 0.3959
    & 0.2994 & 0.4536 & 0.8394
    \\\cmidrule{2-9}
    & \multirow{2}[0]{*}{$\mt$}
    & \NCs\ $\Ala$
    & {0.4774} & {0.8045} & {0.3534}
    & {0.3129} & {0.5451} & {0.7778}
    \\ 
    & & \NCs\ $\MHp$
    & 0.4777 & 0.8098 & 0.3407
    & 0.3114 & 0.5379 & 0.7834
    \\\midrule\midrule
    \multirow{4}[4]{*}{$2$} & \multirow{2}[0]{*}{$\MS$}
    & \NCs\ $\Ala$
    & {0.0447} & {0.4158} & {0.9084}
    & {0.2055} & {0.8354} & {0.5099}
    \\ 
    & & \NCs\ $\MHp$
    & 0.0515 & 0.4267 & 0.9029
    & 0.2017 & 0.8298 & 0.5203
    \\\cmidrule{2-9}
    & \multirow{2}[0]{*}{$\mt$}
    & \NCs\ $\Ala$
    & {0.0232} & {0.3906} & {0.9203}
    & {0.1594} & {0.7772} & {0.6088}
    \\
    & & \NCs\ $\MHp$
    & 0.0162 & 0.3796 & 0.9250
    & 0.1623 & 0.7822 & 0.6015
    \\\midrule\midrule
    \multirow{4}[4]{*}{$3$} & \multirow{2}[0]{*}{$\MS$}
    & \NCs\ $\Ala$
    & {0.8723} & {0.4594} & {0.1674}
    & {0.9326} & {0.3251} & {0.1568}
    \\ 
    & & \NCs\ $\MHp$
    & 0.8723 & 0.4595 & 0.1674
    & 0.9326 & 0.3251 & 0.1569
    \\\cmidrule{2-9}
    & \multirow{2}[0]{*}{$\mt$}
    & \NCs\ $\Ala$
    & {0.8784} & {0.4476} & {0.1678}
    & {0.9363} & {0.3144} & {0.1563}
    \\ 
    & & \NCs\ $\MHp$
    & 0.8784 & 0.4474 & 0.1682
    & 0.9363 & 0.3144 & 0.1563
    \\\bottomrule
  \end{tabular}
 \caption{Absolute  values  for  the  mixing matrix  elements  of  the
 \cp-even scalar sector for TP2--5  
when using the \DRbar\ renormalization
of the top/stop sector for either $\Ala$ as input or $\MHp$.
  \label{tab:ZH13Alam}
}
\end{table}


%% file: sections/conclusion.tex
\section{Conclusions}
\label{sec:conclusion}

We have analyzed the predictions for the Higgs-boson masses and mixing
matrices in the NMSSM based on an OS renormalization of the top/scalar
top sector. We compared the  implementation of the results obtained in
this  scheme  in the  codes  \NC\  and  \NFH\ up  to  \order{\alt\als}
(omitting  further MSSM-like  higher-order corrections  implemented in
\NFH).  Differences in  the calculations implemented in  the two codes
arise  from  different  renormalization  prescriptions  and  different
treatments of the electromagnetic and strong coupling constants, which
provide an  indication of  the possible  size of  unknown higher-order
corrections.     Furthermore    genuine     NMSSM    corrections    of
\order{\alt\als} are implemented in \NC,  and from the comparison with
\NFH\ one  can infer the relevance  of these corrections.  As  a final
step,  going beyond  the OS  prescription  in the  top/sector, also  a
comparison with the \DRbar\ renormalization as implemented in \NC\ has
been performed.  Our work complements and extends the results obtained
in ref.~\cite{Staub:2015aea}, where  the Higgs-boson mass calculations
in  different \DRbar\  codes  had  been compared.   In  order to  make
contact with  this analysis,  we employed the  same scenarios  (TP2 --
TP5) as in~\cite{Staub:2015aea}. (The scenarios TP1 and TP6 have found
not to be  useful for our comparison of \NC\  and \NFH.) The scenarios
are  defined  at  the  stop   mass  scale  $\MS$.  Since  diagrammatic
calculations as implemented in \NC\  and \NFH\ are in general designed
to evaluate the Higgs-boson sector for SUSY scales that are not widely
separated from the  weak scale, we also evolved the  TP scenarios down
to the scale  of the top-quark mass.  All Higgs  mass evaluations have
been  done  at  these  two   scales.   At  both  scales  the  original
\DRbar\  parameters have  been converted  to OS  parameters that  were
subsequently used as input for \NC\ and \NFH.

We started with an ``out-of-the-box''  comparison of the two codes and
found large differences of several GeV between the two codes. In order
to disentangle the origin of  the differences we first concentrated on
the one-loop results.  While at  the one-loop level both codes perform
a  complete calculation,  they differ  in the  renormalization of  the
electromagnetic coupling constant~$\alpha$.  The resulting differences
are formally of electroweak two-loop order. For the further comparison
these differences, which  yield an indication of the  possible size of
unknown higher-order corrections  of this type, have  been adjusted by
reparametrizing \NFH\ to the value used  by \NC.  In a second step, in
the  two-loop  \order{\alt\als}  corrections we  adjusted  the  strong
coupling constant~$\als$ in \NC, where $\als^{\DRbar}(Q)$ is employed,
to $\als^{\MSbar}(\mt)$ as  used by \NFH. Although  this difference is
formally only of three-loop order,  this change improved the agreement
between  the  two codes  by  several~GeV  for  the cases  where  large
discrepancies  had  been  observed.    The  remaining  differences  of
\order{0.5\, \gev}  are due  to the genuine  NMSSM corrections  in the
\order{\alt\als} corrections that are implemented  in \NC, but not yet
in   \NFH.   Conversely,   the  corrections   beyond  \order{\alt\als}
implemented in  \NFH, which  are taken  over from  the MSSM  have been
omitted for this comparison, and their numerical impact on the SM-like
Higgs-boson mass  has briefly  been discussed.  In  the final  step we
used the different renormalization schemes of the top/stop sector that
are implemented in \NC\ (but not  in \NFH). We compared the results of
the OS renormalization  (as obtained before) with the  results using a
\DRbar\  renormalization  in  the  top/stop  sector.   Differences  of
\order{1\, \gev} have been found,  which are indicative of theoretical
uncertainties   due   to   unknown  higher-order   corrections.    The
differences in  the choice of  $\als$ on  the other hand  (see above),
lead to a somewhat larger  estimate of the theoretical uncertainty due
to   missing    higher   orders.     In   order   to    make   contact
with~\cite{Staub:2015aea},  we also  analyzed the  differences between
$\MHp$ and $\Ala$ (as used in  that analysis) as \DRbar\ inputs, which
are  two  possible  input  options  in  \NC.   Here  only  very  small
differences for the SM-like Higgs boson have been found.

In this  paper we have  identified the various sources  of differences
between the presented  calculations within an on-shell  scheme for the
top/stop sector  and between  different renormalization  schemes.  The
analyses performed in  this paper yield a better  understanding of the
remaining   theoretical   uncertainties  from   unknown   higher-order
corrections in the  predictions for the Higgs-boson  masses and mixing
matrix elements in the NMSSM.  These  results can now be used to endow
the theoretical predictions for observables  in the NMSSM Higgs sector
with reliable estimates for the remaining uncertainties.